\documentclass[runningheads]{llncs}

\usepackage[utf8]{inputenc}
\usepackage{xspace}
\usepackage{bm}
\usepackage[disable]{todonotes}
\usepackage{apptools}
\usepackage{amsmath}
\usepackage{amssymb}
\usepackage{mathtools}
\usepackage{thm-restate}
\usepackage[font=small,labelfont=small]{subcaption}
\usepackage[font=small,labelfont=bf,labelsep=period]{caption}
\usepackage{xcolor, colortbl}
\usepackage[inline]{enumitem}
\usepackage[misc,geometry]{ifsym}
\usepackage[pdfpagelabels]{hyperref}
\usepackage[capitalise]{cleveref}
\usepackage{orcidlink}

\newcommand{\arxivlncs}[2]{#1} %

\newcommand{\restateref}[1]{\arxivlncs{\IfAppendix{\hyperref[#1]{$\star$}}{\hyperref[#1*]{$\star$}}}{\href{http://arxiv.org/pdf/XXXX.YYYYY}{$\star$}}}

\graphicspath{{figures/}}

\definecolor{lightcyan}{rgb}{0.88,1,1}
\definecolor{antiquewhite}{rgb}{0.98, 0.92, 0.84}
\definecolor{lightgreen}{rgb}{0.6, 0.98, 0.6}
\definecolor{defblue}{rgb}{0.121,0.47,0.705}

\DeclareTextFontCommand{\emph}{\color{defblue}\em}

\newcommand{\D}{\ensuremath{\mathcal D}\xspace}
\newcommand{\W}{\ensuremath{\mathcal W}\xspace}
\newcommand{\LL}{\ensuremath{\mathcal L}\xspace}
\newcommand{\WL}{\ensuremath{\mathcal{W}_{\overline{\mathcal{L}}}}\xspace}
\newcommand{\C}{\ensuremath{\mathcal{C}}\xspace}
\newcommand{\CC}{\ensuremath{\mathcal{C}^{+}}\xspace}

\title{Drawings of Trees and Outerplanar Graphs with~Few Directions of Monotonicity}

\titlerunning{Drawings with Few Directions of Monotonicity}

\author{Eleni Katsanou\inst{1}\orcidlink{0000-0002-1001-1411} and
	  Alexander Wolff\hspace*{.2ex}\inst{2}\orcidlink{0000-0001-5872-718X}
	}%
	
\institute{National Technical University of Athens, Athens, Greece \\
		  \email {ekatsanou@mail.ntua.gr}
		  \and
		  Universit\"at W\"urzburg, W\"urzburg, Germany
		}%

\authorrunning{E.~Katsanou and A.~Wolff}

\let\doendproof\endproof
\renewcommand\endproof{~\hfill$\qed$\doendproof}

\begin{document}

\maketitle
		
\begin{abstract}
  A drawing of a path is \emph{monotone with respect to a line~$l$} if
  the orthogonal projections of its vertices onto~$l$ appear strictly
  in the same order along~$l$ as they appear on the path.  A drawing
  of a graph~$G$ is \emph{monotone} if, for every pair of vertices
  in~$G$, there exists a path connecting them that is monotone with
  respect to some line.  We call a drawing of a graph
  \emph{$k$-monotone} if it is monotone with respect to a set of $k$
  lines.
  In this paper, we investigate the number of lines
  required to construct monotone straight-line drawings of trees and
  outerplanar graphs.

  We present an algorithm that, for every tree with $\ell$ leaves,
  constructs an $\lceil \ell/2 \rceil$-monotone drawing (where all
  edges have unit length).  Since it is known that trees
  with $\ell$ leaves do not admit drawings with less than
  $\lceil \ell/2 \rceil$ lines of monotonicity, our result is tight.
  All existing approaches for monotone drawings of trees produce
  drawings that are monotone with respect to at least $\ell$ lines.
  
  Our main result is that every outerplanar graph~$G$ admits a
  $\max\{2,\allowbreak\lceil \ell(G)/2 \rceil\}$-monotone drawing, where \emph{$\ell(G)$} is the number of
  leaves of the block-cut tree of~$G$, except if~$\ell(G) = 4$ and some block of~$G$ contains at least three cut vertices; then~$G$ admits a $3$-monotone drawing.  We show that these bounds are
  tight.  Prior techniques rely on constructing a monotone
  drawing of a spanning tree, which in many cases
  has linearly more leaves than the block-cut tree.

\end{abstract}

\section{Introduction}
\label{se:introduction}
	
	A plane path is \emph{monotone with respect to a line~$l$} if the orthogonal
	projections of its vertices onto~$l$ appear strictly in the same order along~$l$ as they appear on the path. Any
	monotone path is necessarily crossing-free~\cite{AngeliniCBFP12}. A drawing of a graph~$G$
	is \emph{monotone} if, for every pair of vertices in~$G$, there exists a path connecting
	them that is monotone with respect to some line. When the lines of monotonicity are restricted to a
	prescribed set~$\LL$ of lines through the origin, the drawing is called \emph{$\LL$-monotone}.
        If we are just interested in the number $k=|\LL|$ of lines of monotonicity, we can call the drawing also \emph{$k$-monotone}.
        Correspondingly, we call a graph \emph{$k$-monotone} if it admits a $k$-monotone drawing.
	
	Monotone drawings of graphs were introduced by Angelini,
	Colasante, {Di Battista}, Frati, and Patrignani~\cite{AngeliniCBFP12}; such drawings have received considerable attention in recent years.
	They are related to other well-studied drawing paradigms, such as
	upward drawings~\cite{DBLP:reference/algo/Didimo16,gt-upt-95} and
	greedy drawings~\cite{DBLP:journals/tcs/AngeliniBDGKMPS19,angelini10,dehkordi14,Dhandapani2010,Papadimitriou2005,Rao2003},
	as well as the geometric problem of finding monotone trajectories
	between two points in the plane amidst convex obstacles~\cite{DBLP:conf/compgeom/ArkinCM89}.
	
	A natural focus is on planar monotone drawings with straight-line edges.
	Angelini, Colasante, {Di Battista}, Frati, and Patrignani~\cite{AngeliniCBFP12} showed that trees and biconnected planar graphs always admit such drawings, and Hossain and Rahman~\cite{hossainR15} extended this to all connected planar graphs.
	Monotone drawings have also been investigated in the fixed-embedding setting~\cite{angeliniDKMRSW15} and under stronger requirements, such as strongly monotone drawings of planar graphs~\cite{AngeliniCBFP12,kindermannSSW14,felsnerIKKMS16,NollenburgPR14}.
	From a geometric perspective, the problem has also been examined within the scope of computing a minimum monotone spanning tree for a given planar point set and a prescribed set of lines~\cite{ddkssw-mmst-algorithmica}.
	While the area requirements of monotone drawings have been studied extensively for trees~\cite{hehe17,kindermannSSW14,oikonomouSym18} and other planar graph classes~\cite{hehe15b,hossainR15,oikonomouSymv17}, the number of lines (or directions) required to realize them remains largely unexplored.
	Angelini~\cite{Angelini17} showed that a planar graph is 1-monotone if and only  if it contains a Hamiltonian path whose endpoints lie on a common face.
	Furthermore, using
	classical Schnyder drawings~\cite{Schnyder90}, Angelini showed that
	maximal planar graphs are 2-monotone and general 3-connected planar graphs are 3-monotone.
	Whether every $3$-connected planar graph is 2-monotone remains an open question.

    \paragraph{Our Contribution.}
	In this paper, we investigate the number of lines of monotonicity required to construct monotone drawings of trees and outerplanar graphs.
	
	For trees~\cite{hehe17,kindermannSSW14,oikonomouSym18}, all existing algorithms in the literature produce drawings that are monotone with respect to at least $\ell$ lines, where $\ell$ is the number of leaves of the given tree.
        On the other hand, {Di Giacomo}, Didimo, Katsanou, Schlipf, Symvonis, and Wolff~\cite{ddkssw-mmst-algorithmica} showed that $\lceil \ell/2 \rceil$ lines are necessary.
        We close this gap by presenting an algorithm that, for every tree with $\ell$ leaves, constructs an $\lceil \ell/2 \rceil$-monotone drawing; see \cref{thm:trees} in \cref{sec:trees}.
        
	For outerplanar graphs, we show that every outerplanar graph~$G$ admits a~$\max\{2,\lceil \ell(G)/2 \rceil\}$-monotone drawing, where~\emph{$\ell(G)$} is the number of leaves of the block-cut tree of~$G$ (defined at the end of \cref{se:preliminaries}), except if~$\ell(G) = 4$ and some block of~$G$ contains at least three cut vertices; then~$G$ admits a $3$-monotone drawing.  This is our main result; see \cref{thm:upperbound} in \cref{sec:cacti}.
	Furthermore, we show that our result is tight; for every~$\ell \ge 2$, there is an outerplanar graph~$G$ with~$\ell(G)=\ell$ in which every block contains at most two cut vertices such that any monotone drawing of~$G$ requires at least~$\max\{2,\lceil \ell/2 \rceil\}$ lines of monotonicity, and there is an outerplanar graph~$G$ with~$\ell(G) = 4$ that requires three lines in every planar monotone drawing; see \cref{thm:lowerbound}.
	\begin{figure}[b]
		\centering
		\includegraphics[page=4]{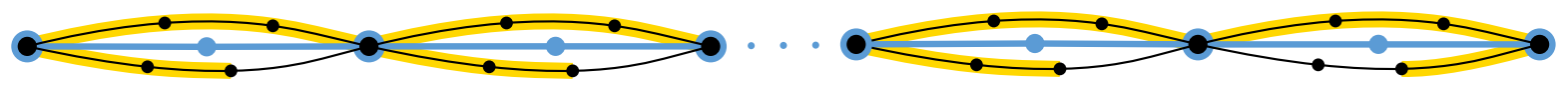}
		\caption{A graph $G$ (in black) with $f = \Theta(|V(G)|)$ faces, its block-cut tree $T_G$ (in blue) with two leaves, and a spanning tree of $G$ (in yellow) with $f$ leaves.}
		\label{fig:leaves}
	\end{figure}
	Prior techniques \cite{angeliniDKMRSW15,oikonomouSymv17} rely on constructing a monotone drawing of a spanning tree of the given graph.
	However, any spanning tree of an outerplanar graph has at least as many leaves as its block-cut tree.
        \Cref{fig:leaves} shows an outerplanar graph~$G$ whose
        block-cut tree has two leaves, whereas every spanning tree
        of~$G$ has a linear number of leaves.
        
    \arxivlncs{The proofs of statements with a (clickable) ``$\star$''
    appear in the appendix.}{Missing proofs (marked ``$\star$'') can be found in the full version of this paper~\cite{katsanouW-arxiv26}.}

	\section{Basic Definitions and Properties}
	\label{se:preliminaries}
	
\begin{figure}[tb]
		\captionsetup[subfigure]{justification=centering}
		\begin{subfigure}{.48\linewidth}
			\centering
			\includegraphics[page=4]{path_sector_wedges}
			\subcaption{}
			\label{fig:lines_directions_1}
		\end{subfigure}
		\hfill
		\begin{subfigure}{.48\linewidth}
			\centering
			\includegraphics[page=5]{path_sector_wedges}
			\subcaption{}
			\label{fig:lines_directions_2}
		\end{subfigure}
		\hfill
		\begin{subfigure}{.31\linewidth}
			\centering
			\includegraphics[page=1]{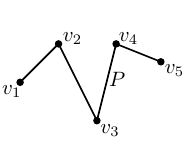}
			\subcaption{}
			\label{fig:path_sector_wedges_1}
		\end{subfigure}
		\hfill
		\begin{subfigure}{.31\linewidth}
			\centering
			\includegraphics[page=2]{path_sector_wedges}
			\subcaption{}
			\label{fig:path_sector_wedges_2}
		\end{subfigure}
		\hfill
		\begin{subfigure}{.31\linewidth}
			\centering
			\includegraphics[page=3]{path_sector_wedges}
			\subcaption{}
			\label{fig:path_sector_wedges_3}
		\end{subfigure}
		\caption{(a)~A set~$\LL=\{l_0, \dots, l_{4}\}$ of five lines and the corresponding set of directions~$\D_\LL=\{d_0, \dots, d_9\}$.
			(b)~The wedge set defined by the lines in~$\overline{\LL}$.
			(c)~A directed geometric path~$P$. (d)~The
                  sector of directions of~$P$,~$\sec(P)$ (in dark
                  gray). (e)~The set of (four) wedges~$\W_P$ (medium gray) utilized
                  by~$P$, with respect to the set of
                  lines~\LL from~(a).}
		\label{fig:path_sector_wedges}
	\end{figure}

	A \emph{geometric graph} is a graph whose vertices are points in the plane and whose edges are straight-line segments between them; in particular, a straight-line drawing of a graph is a geometric graph. Geometric paths and trees are defined accordingly.
		A \emph{direction} is defined as a unit vector 
	from the origin to a point on the unit circle.
	A geometric path~$P = \langle v_1, v_2, \dots, v_n \rangle$ is \emph{$d$-monotone} if the order of its vertices along the path coincides with the order of their projections on any directed line parallel to~$d$ and directed as~$d$.
	Note that, in this case, the path~$\langle v_n, \dots, v_1 \rangle$ is~$(-d)$-monotone.
	A path is called \emph{monotone} if it is~$d$-monotone for some direction~$d$.
	Let~$\LL=\{l_0, \dots, l_{k-1}\}$ be a set of~$k$ lines through the origin. Refer to \cref{fig:lines_directions_1}. We assume that the lines in \LL are numbered as~$l_0, \dots l_{k-1}$ in the order in which we encounter their directions when rotating the vector~$(1,0)$ counterclockwise. 
	Let \emph{$\D_\LL$}~$=\{d_0, \dots, d_{2k-1}\}$ be the set of the~$2k$ directions defined by the lines in~\LL, where each line defines two opposite directions.
	We assume that the vectors in~$\D_\LL$ are numbered in the order in which we encounter them when rotating the vector~$(1,0)$ counterclockwise. 
	Note that, for every~$i \in \{0, \dots, k-1\}$,~$d_{i+k}=-d_i$ and that both vectors lie on~$l_i$.
	A path is called~$l_i$-monotone if it is~$d_i$- or~$d_{i+k}$-monotone.
	A drawing of a graph~$G$ in the plane is \emph{\LL-monotone} if, for every pair of vertices~$\{u, v\}$ in~$G$, there exists a path between~$u$ and~$v$ that is~$l$-monotone for a line~$l \in \LL$. A graph~$G$ is \emph{\LL-monotone} if it admits an \LL-monotone drawing. 
	Given a line~$l$ and a point~$p$, let \emph{$l(p)$} denote the line parallel to~$l$ passing through~$p$, and let \emph{$\overline{l}(p)$} denote the line orthogonal to~$l(p)$ in~$p$. 
	Given a direction~$d$, let~$\overline{d}$ denote the direction orthogonal to~$d$, obtained by rotating~$d$ counterclockwise by~$90^\circ$.

	For any point~$p$ in the plane, the lines \emph{$\overline{\LL}(p)$}~$=\{\overline{l_0}(p), \overline{l_1}(p), \dots, \overline{l_{k-1}}(p)\}$ partition the plane into~$2k$ wedges with apex~$p$.
        Let \emph{$\WL(p)$}~$= \{W_0(p), \dots, W_{2k-1}(p)\}$ be this set of wedges, where, for every~$i \in \{0, \dots, 2k-1\}$, wedge~$W_i(p)$ is delimited by the
        directed half-lines defined by~$\overline{d_i}$ and~$\overline{d_{(i+1) \bmod 2k}}$
        originating in~$p$. Refer to \cref{fig:lines_directions_2}.
	The index~$i$ in~$W_i(p)$ is always interpreted modulo~$2k$. When the reference point~$p$ coincides with the origin~$o$, we simplify notation by writing~\mbox{\emph{$\overline{\LL}$}~$=\{\overline{l_0}, \overline{l_1}, \dots, \overline{l_{k-1}}\}$} instead of~$\overline{\LL}(o)$ and
        \emph{$\WL$}~$= \{W_0, W_1, \dots, W_{2k-1}\}$
        instead of~$\WL(o)$.
	
	Consider a directed geometric path~$P=\langle v_1,\dots,v_n \rangle$;
    see \cref{fig:path_sector_wedges_1}.
    For~$i \in \{1,\dots,n-1\}$, let~$\overrightarrow{\sigma_i}$
    be the vector that is parallel to and has the same orientation as~$\overrightarrow{v_i v_{i+1}}$.
	Define the \emph{sector of directions} of path~$P$, denoted by
	\emph{$\sec(P)$}, to be the smallest sector of the unit circle that
        includes all vectors~$\overrightarrow{\sigma_i}$,~$1\leq i <n$; see \cref{fig:path_sector_wedges_2}. 
	Given a finite set~\LL of lines through the origin,
        let \emph{$\W_{P}$}~$\subseteq \WL$
        be the smallest set of consecutive
	wedges (in counterclockwise order) whose union contains~$\sec(P)$.
    We say that path~$P$ \emph{utilizes} wedge
        set~$\W_{P}$; see \cref{fig:path_sector_wedges_3}.
        Finally, for a point~$p$ in the plane, we define~$\W_P(p)$
        to be the translation of~$\W_P$ so that~$p$ is its apex.
        We let~$\W_P^{-}$ denote the set consisting of the wedges opposite to those in~$\W_P$.  
        Clearly, if the directed path~$P=\langle v_1,\dots,v_n\rangle$
        utilizes the wedge set~$\W_P$, then the reverse
	path~$\langle v_n,\dots,v_1\rangle$ utilizes the wedge set~$\W_P^{-}$.

	We use the following properties of monotone paths established
        in~\cite{ddkssw-mmst-algorithmica}.
	
	\begin{lemma}[\cite{ddkssw-mmst-algorithmica}]
          \label{lem:max-degree-2k}
          Let \LL be a set of~$k$
          lines through the origin. If~$T$ is an \LL-monotone tree, then~$T$ has at
          most~$2k$ leaves and~$\Delta(T) \leq 2k$.
	\end{lemma}
	
	\begin{lemma}[\cite{ddkssw-mmst-algorithmica}]
		\label{lem:separating_dir}
		Let~\LL be a set of lines through the origin, and let~$P$
                be a path that contains vertices~$u$, $x$, $v$ in that order.
		If~$u$ and~$v$ lie in the same wedge in~$\WL(x)$, then
		path~$P$ is not \LL-monotone. 
	\end{lemma}
	
	\begin{lemma}[\cite{ddkssw-mmst-algorithmica}]
          \label{lem:monot_external_dir}
          Given a line~$l$ through the origin and a path~$P$, then $P$
          is~$l$-monotone if and only if~$\overline{l}$ does not
          intersect the interior of~$\W_P$.
	\end{lemma}
	
	We assume that no edge is parallel to a line of~$\overline{\LL}$; otherwise, \cref{lem:monot_external_dir} fails, since an edge parallel to~$\overline{l}$ is not $l$-monotone, although~$\overline{l}$ does not intersect the interior of a wedge that contains this edge. 
		
	Note that a direct implication of
        \cref{lem:monot_external_dir} is that, if a path~$P$ is
        $\LL$-monotone, then~\mbox{$|\W_{P}| \le |\LL|$}. We
        frequently leverage this property to establish the
        $\LL$-monoto\-ni\-city of paths.  Using the above lemmas, we
        show the following result.
	
	\begin{restatable}[\restateref{lem:wedges_disjoint}]{lemma}{wedgesDisjoint}
			\label{lem:wedges_disjoint}
			Let~$\LL$ be a set of lines through the origin, let~$u$,~$x$, and~$v$
			be three vertices such that the path~$\Pi$ from~$u$ to~$v$ passes through~$x$. Let~$P$ be the oriented subpath of~$\Pi$ from~$x$ to~$u$, and let~$Q$ be the oriented subpath of~$\Pi$ from~$x$ to~$v$. If both~$P$ and~$Q$ are~$\LL$-monotone and their wedge sets~$\W_{P}$ and~$\W_{Q}$ are disjoint, then~$\Pi$ is~$\LL$-monotone.
	\end{restatable}
	
	For a set~$\W$ of at most~$|\LL|$ consecutive wedges, let~\emph{$\LL(\W)$}~$\subseteq \LL$ be the set of lines~$l$ such that~$\overline{l}$ does not intersect the interior of the union of the wedges of~$\W$; by \cref{lem:monot_external_dir}, a path that utilizes wedges in~$\W$ is monotone with respect to every line of~$\LL(\W)$.
	For~$l \in \LL(\W)$, the wedges of~$\W$ lie on one side of~$\overline{l}$, and we let~$d_l$ be the direction of~$l$ that points to this side. 
	Let~\emph{$\D(\W)$}~$=\{d_l \mid l \in \LL(\W)\}$, then, 
	a path that utilizes wedges in~$\W$ is $d$-monotone for every~$d \in \D(\W)$.
	We say that a vertex~$v$ is \emph{reachable from~$x$ within~$\W$} if, for every~$d \in \D(\W)$, there is a $d$-monotone path from~$x$ to~$v$.
	We say that~$v$ is \emph{weakly reachable from~$x$ within~$\W$} if there exists a direction~$d \in \D(\W)$ such that there is a $d$-monotone path from~$x$ to~$v$.
	The following lemma generalizes \cref{lem:wedges_disjoint}.
	
	\begin{restatable}[\restateref{lem:reach_disjoint}]{lemma}{reachDisjoint}
			\label{lem:reach_disjoint}
			Let~$\W$ and~$\W'$ be disjoint sets of at most~$|\LL|$ consecutive wedges of~\WL each, and let~$u$ and~$v$ be reachable from a vertex~$x$ within~$\W$ and within~$\W'$, respectively. Then~$u$ and~$v$ are connected by an \LL-monotone path.
	\end{restatable}
	
	The following lemma provides a key building block for our proofs by showing how to construct the geometric paths that form our drawings. 
	
	\begin{restatable}[\restateref{lem:convex_path}]{lemma}{convexPath}
          \label{lem:convex_path}
          Let~$\LL$ be a set of lines through the origin, let~\W be a
          set of at most~$|\LL|$ consecutive wedges of~\WL, and let~$W$ be the union
          of the wedges in~\W.  Given two points~$x$ and~$y$
          with~$y$ in the interior of~$W(x)$ and a positive integer~$n$, we can place
          points $p_1, \dots, p_{n}$ such that
          \begin{enumerate*}[label=(\roman*)]
          \item the path~$\langle x, p_1, \dots, p_{n}, y \rangle$ utilizes only wedges in~$\W$ and
          \item the polygon~$\langle x, p_1, \dots, p_{n}, y, x \rangle$ is convex.
          \end{enumerate*}
          Moreover, we can choose on which side of the line through~$x$ and~$y$ the points~$p_1, \dots, p_n$ lie.
	\end{restatable}
	
	The \emph{block-cut tree} of a connected graph~$G$, denoted by~$T_G$, has a vertex for each block (biconnected component) and for each cut-vertex of~$G$. An edge connects a block vertex~$B$ and a cut-vertex~$c$ in~$T_G$ if and only if~$c$ belongs to~$B$.
        
	\section{Trees}
    \label{sec:trees}
	In this section, we show that for any tree~$T$ with~$\ell$ leaves, $\lceil \ell/2 \rceil$ lines are both necessary and sufficient to construct an~$\LL$-monotone drawing. Here, as in the rest of the paper, we assume that~$T$ comes with a fixed \emph{combinatorial embedding}, that is, a fixed cyclic order of the edges around each vertex, and we require that the drawing respects this embedding.
	In a tree, a \emph{branching vertex} is
	a vertex of degree at least~three, and a \emph{leaf path} is a path of
degree-2 vertices from a branching vertex to a leaf.  A \emph{leaf centroid} of~$T$ is a branching
	vertex~$v$ such that every component of~$T - v$
	contains at most~$\ell/2$ leaves of~$T$.
	If~$\ell \ge 3$, then~$T$ has a leaf centroid, which is not necessarily unique.
        If, for two branching vertices~$u$ and~$v$, the unique~$u$--$v$ path
	$B_{u,v}$ does not contain any other branching vertices, then
        we say that~$B_{u,v}$ is a \emph{branch}.  Both a leaf path and a
	branch may consist of a single edge. Further, for any pair of
        (not necessarily adjacent) vertices~$u$ and~$v$, let~$T_{u,v}$
        be the subtree of~$T$ consisting of vertex~$u$ and all
        subtrees hanging from~$u$ except the one containing vertex~$v$.
	Moreover, for a finite set \LL of lines through the origin,
	let
	$\W_{u,v} \subseteq \WL$ be the smallest set of consecutive
	wedges that contains all wedges utilized by either leaf paths or
	branches oriented away from~$u$ in~$T_{u,v}$.
	
\begin{figure}[tb]
		\centering
		\begin{subfigure}[b]{.34\linewidth}
			\centering
			\includegraphics[page=1]{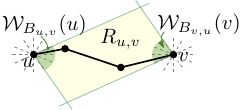}
			\subcaption{$|\W_{B_{u,v}}|<|\LL|$}
			\label{fig:path_regions_1}
		\end{subfigure}
		\qquad
		\begin{subfigure}[b]{.34\linewidth}
			\centering
			\includegraphics[page=2]{path_regions}
			\subcaption{$|\W_{B_{u,v}}|=|\LL|$}
			\label{fig:path_regions_2}
		\end{subfigure}
		\caption{The region~$R_{u,v}$ (yellow) of a branch~$B_{u,v}$ depending on~$|\W_{B_{u,v}}|$ and~$|\LL|$.}
		\label{fig:path_regions}
	\end{figure}

	Let~$B_{u,v}$ be a branch of a monotone tree~$T$ connecting branching
	vertices~$u$ and~$v$, and let
	$R_{u,v}=\W_{B_{u,v}}(u)\cap \W_{B_{v,u}}(v)$; see \cref{fig:path_regions}.
	If~$|\W_{B_{u,v}}|<|\LL|$, then~$R_{u,v}$ is a parallelogram.
	Otherwise, if~$|\W_{B_{u,v}}|=|\LL|$, then the path~$B_{u,v}$ utilizes exactly
	half of the wedges of~$\WL$ and, by \cref{lem:monot_external_dir},~$B_{u,v}$ is
	monotone only with respect to the line~$l \in \LL$ whose orthogonal~$\overline{l}$ delimits
	$\W_{B_{u,v}}$.
	In this case,~$R_{u,v}$ is a strip bounded by the parallel lines
	$\overline{l}(u)$ and~$\overline{l}(v)$.
	We refer to~$R_{u,v}$ as the
	\emph{region of the branch}~$B_{u,v}$.  Similarly, given a leaf
        path~$P_{u, \lambda}$ that goes from a branching vertex~$u$ to
        a leaf~$\lambda$, we define~$R_{u, \lambda}=\W_{P_{u, \lambda}}(u) \cap \W_{P_{\lambda, u}}(\lambda)$
        to be the \emph{region of the leaf path}~$P_{u, \lambda}$.
	Di Giacomo et al.~\cite{ddkssw-mmst-algorithmica} characterized \LL-monotone trees. 

	\begin{theorem}[\cite{ddkssw-mmst-algorithmica}]
		\label{thm:trees_characterization}
		Let~$\LL$ be a finite set of lines through the origin, and let~$T$ be a geometric tree.  Then~$T$ is
		$\LL$-monotone if and only if the following conditions hold.
		\begin{enumerate*}[label=(\alph*)]
			\item Every leaf path and every branch~$P$ in~$T$ is~$\LL$-monotone.
			\item For every two leaf paths~$P$ and~$Q$ incident to branching vertices~$u$ and~$v$, respectively,~$\W_P$ and~$\W_Q$ are disjoint.
			\item For every branch or leaf path from~$u$ to~$v$ in~$T$, the interiors of~$\W_{u,v}(u)$ and~$R_{u,v}$ are disjoint. 
		\end{enumerate*}
	\end{theorem}
	
	The following result follows rather directly from
        \cref{thm:trees_characterization}.
        In particular for the upper bound, we only need to assign the
        wedges to the leaves of~$T$ such that the conditions of
        \cref{thm:trees_characterization} are met.  We sketch the
        proof here since our construction for outerplanar graphs in
        \cref{sec:cacti} builds on it.
	
	\begin{restatable}[\restateref{thm:trees}]{theorem}{trees}
          \label{thm:trees}
          Let~$T$ be a tree with~$\ell$ leaves and a fixed combinatorial embedding, and let~$\ell'=\lceil \ell/2 \rceil$. If~\LL is a set
          of~$\ell'$ lines, then~$T$
          admits an~$\LL$-monotone drawing. 
          On the other hand, for any set~$\LL$ of lines such that~$T$ admits an~$\LL$-monotone drawing, it holds that~$|\LL| \geq \ell'$.
	\end{restatable}

	\begin{proof}[sketch]
          The lower bound follows from \cref{lem:max-degree-2k}.  For
          the upper bound, let $u$ be a leaf centroid of~$T$, root~$T$
          at~$u$, and assign the wedges~$W_0, \dots, W_{2\ell'-1}$ in
          counterclockwise order to the leaves in DFS-order, following
          the embedding of~$T$.  Assign one wedge per leaf and two
          wedges to the last leaf if~$\ell$ is odd.  Thus, every
          subtree below~$u$ receives a set of at most~$\ell'$
          consecutive wedges.
          
		We draw every branch and leaf path with unit-length edges along the bisector of the wedges assigned to the leaves below it and verify conditions~(a)--(c) of \cref{thm:trees_characterization}; see \cref{fig:tree_branch_regions_3} for an example.
	\end{proof}

    \section{Outerplanar graphs}
    \label{sec:cacti}
    
    In this section, we show the following two main results.
    
    First, in \cref{thm:upperbound}, we demonstrate that every outerplanar
    graph~$G$ with a fixed outerplanar embedding admits, for every set~\LL
    of~$\max\{2,\lceil \ell(G)/2 \rceil\}$ lines, an \LL-monotone drawing
    with convex interior faces, except if~$\ell(G) = 4$ and some block of~$G$ contains at least three cut vertices, in which case three lines suffice.  Given such an embedded outerplane graph,
    a \emph{chord} is an internal edge that connects two non-consecutive
    vertices along the boundary of a bounded face of the graph.  Note
    that, if~$G$ is an outerplanar graph, then by deleting all chords
    from~$G$, we obtain a spanning subgraph~$G'$ of~$G$.  This subgraph is
    a \emph{cactus graph} (or \emph{cactus}), that is, a connected graph
    in which every edge lies on at most one cycle; equivalently, every
    block of it is an edge or a cycle.  Every cactus is outerplanar.
    Thus, to show \cref{thm:upperbound}, it suffices to draw the
    cactus~$G'$ with convex interior faces and then to reinsert the
    deleted chords.
    Second, in \cref{thm:lowerbound}, we prove that this bound is worst-case optimal. For every~$\ell \ge 2$, some cactus~$G$ with~$\ell(G)=\ell$ in which every block contains at most two cut vertices requires at least~$\max\{2,\lceil \ell/2 \rceil\}$ lines of monotonicity, and some cactus~$G$ with~$\ell(G) = 4$ requires three lines in every planar monotone drawing.
    
    We start by giving some auxiliary notation and definitions. 
    Let~$G$ be a cactus graph with a fixed outerplanar embedding~$\mathcal E$.
    In this case, the blocks of $G$ are exactly the bounded faces and the edges that do not lie on any bounded face.
	
	Let~$c^\star$ be a cut vertex of~$G$ corresponding to node~$v_{c^\star}$ of~$T_G$. We root~$T_G$ at~$v_{c^\star}$ and orient its edges away from it.  
	For a node~$v_f$ in~$T_G$ corresponding to a bounded face~$f$ of~$G$, we define
	the \emph{parent vertex}~$c$ of~$f$ as the cut vertex of~$G$ corresponding to the parent node of~$v_f$ in~$T_G$, and
	the \emph{child vertices}~$c_1, c_2, \dots, c_m$ of~$f$ as the cut vertices of~$G$ (in counterclockwise order) corresponding to the child nodes of~$v_f$ in~$T_G$. We let~$\C=\{c_1, c_2, \dots, c_m\}$ denote the set of child vertices of~$f$, and by~$\CC = \C \cup \{c\}$ the set of cut vertices of~$f$.
	\begin{figure}[tb]
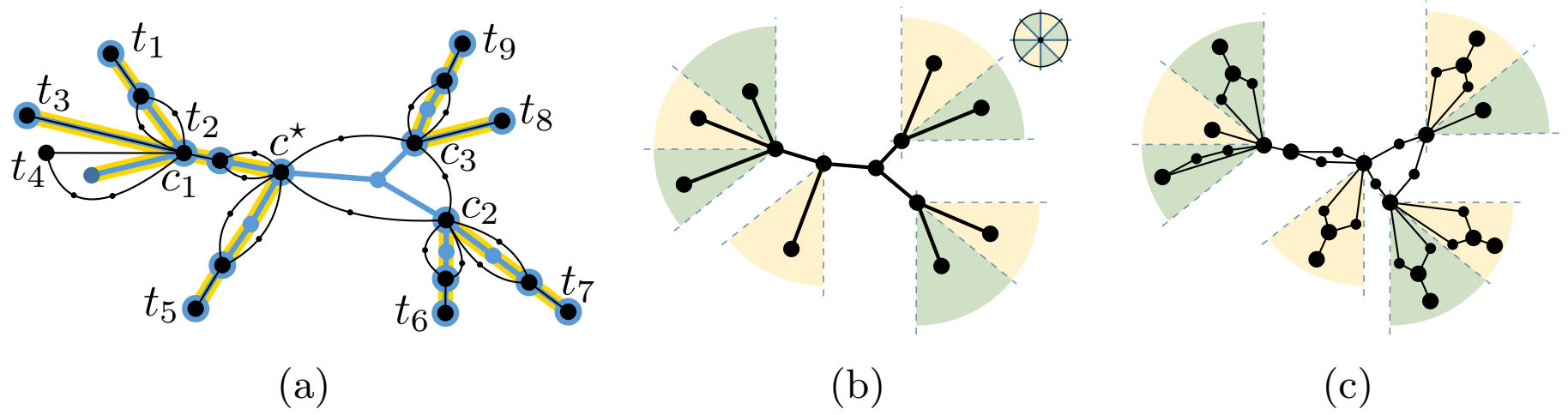

		\captionsetup[subfigure]{justification=centering}
		\begin{subfigure}{.38\textwidth}
			\centering
			\includegraphics[page=2]{bc_tree}
			\subcaption{}
			\label{fi:bc_tree_1}
		\end{subfigure}
		\hfill
		\begin{subfigure}{.28\textwidth}
			\centering
			\includegraphics[page=3]{bc_tree}
			\subcaption{}
			\label{fig:tree_branch_regions_3}
		\end{subfigure}
		\hfill
		\begin{subfigure}{.3\textwidth}
			\centering
			\includegraphics[page=4]{bc_tree}
			\subcaption{}
			\label{fi:bc_tree_2}
		\end{subfigure}
		\caption{(a) A cactus graph~$G$ (in black) with its
                  block-cut tree~$T_G$ (in blue) rooted
                  at~$v_{c^\star}$ and its maximal paths (in
                  yellow). Vertices~$c^\star$, $c_1$, $c_2$, and $c_3$
                  are the roots of the maximal chains, and
                  vertices~$t_1,\dots,t_9$ are the corresponding
                  terminal vertices.  (b)~An \LL-monotone drawing of
                  the tree~$T_G$.  (c)~An \LL-monotone drawing
                  of~$G$.}
		\label{fi:bc_tree}
	\end{figure}
	For any node~$v_{c}$ in~$T_G$ corresponding to a cut vertex~$c$, let~$T_{v_{c}}$ be the subtree of~$T_G$ rooted at~$v_{c}$. 
	Let~$G_{c}$ be the subgraph of~$G$ induced by the vertices and components corresponding to the nodes of~$T_{v_{c}}$.
	
	Now, let~$P = \langle v_0, v_1, \dots, v_m \rangle$ be a path in~$T_G$ that is directed away from the root, where~$v_0$ corresponds to a cut vertex~$c_0$ of~$G$ and each intermediate vertex~$v_i$ (for~$1 \le i < m$) has degree two in~$T_G$. 
	The terminal vertex~$v_m$ is either a leaf of~$T_G$ or corresponds to another cut vertex of~$G$. We refer to a subgraph of~$G$ that corresponds to such a path as a \emph{chain}.
	If~$P$ cannot be extended to a larger path while maintaining these properties, neither at~$v_0$ nor at~$v_m$, then path~$P$ is called a \emph{maximal path} of~$T_G$ and the corresponding subgraph of~$G$ is called a \emph{maximal chain}; see \cref{fi:bc_tree}.
	We refer to~$c_0$ as the \emph{root} of the chain. 
	If~$v_m$ corresponds to a cut vertex or a degree-one vertex~$c_m$ in~$G$, we refer to~$c_m$ as the \emph{terminal vertex} of the chain. 
	If~$v_m$ corresponds to a face~$f$ with parent vertex~$c_i$, we designate an arbitrary neighbor of~$c_i$ in~$f$ as the terminal vertex.
	Observe that~$G$ is composed of maximal chains and of \emph{high-degree faces}, which are faces containing at least three cut vertices.
	We refer to maximal chains and high-degree faces as the \emph{components} of~$G$.

	Our construction for \cref{thm:upperbound} builds on the characterization of \LL-monotone trees by Di Giacomo et al.~\cite{ddkssw-mmst-algorithmica} (\cref{thm:trees_characterization}), applied to the block-cut tree~$T_G$ instead of a tree. The maximal chains and the high-degree faces of~$G$ take the role of leaf paths and branches, respectively, and we assign wedges to the leaves of~$T_G$ as in the proof of \cref{thm:trees}.
	The three conditions of \cref{thm:trees_characterization} roughly correspond to the requirements of our construction. Every two vertices of the same chain or face are connected by an \LL-monotone path~(a), the components that hang from a cut vertex are reachable within pairwise disjoint sets of consecutive wedges~(b), and every component is drawn inside the region defined by its own wedges~(c).
	The main difference is that the blocks of~$G$ may be faces whose two sides cannot both utilize a fixed set of wedges; hence, we relax the property that a path utilizes a set of wedges to the property that its endpoint is reachable within this set.

	Next, we describe how to draw maximal chains.
        
	\begin{lemma}\label{lem:face-1-wedge}
		Let~$\W \subset \WL$ be a set of at most~$|\LL|$ consecutive wedges, and let~$f$ be a bounded face of~$G$.
		If~$f$ is incident to a single cut vertex~$c$, then~$f$ admits a
		convex drawing with~$c$ at a given position such that, for the counterclockwise boundary~$\langle c = v_0, v_1, \dots, v_m \rangle$ of~$f$ and for~$0 \le i < j \le m$, the path~$\langle v_i, \dots, v_j \rangle$
			utilizes wedges in~$\W$ and, hence, is \LL-monotone. In particular, every vertex of~$f$ other than~$c$ lies in the interior of the region~$\W(c)$.
	\end{lemma}
	
\begin{figure}[tb]
		\captionsetup[subfigure]{justification=centering}
		\begin{subfigure}{.45\linewidth}
			\centering
			\includegraphics[page=2]{face_monotonicity}
			\subcaption{}
			\label{fig:face_monotonicity_1}
		\end{subfigure}
		\hfill
		\begin{subfigure}{.45\linewidth}
			\centering
			\includegraphics[page=1]{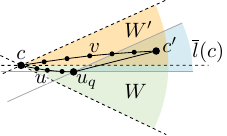}
			\subcaption{}
			\label{fig:face_monotonicity_2}
		\end{subfigure}
		\caption{(a)~The construction of a face~$f$ with no child vertex. (b)~The construction of a face~$f$ with one child vertex.}
		\label{fi:face_monotonicity}
	\end{figure}

	\begin{proof}
		Let~$W = \bigcup \W$ be the smallest wedge that contains the wedges in~$\W$, and let~$\langle c, v_1, \dots, v_m \rangle$ be the counterclockwise boundary of~$f$ with cut-vertex~$c$ (see \cref{fig:face_monotonicity_1}). 
		We place~$c$ at its given position and~$v_m$ along the angular bisector of~$W(c)$, rotated slightly if this bisector lies on a line of~$\overline{\LL}$.
		By applying \cref{lem:convex_path} between~$c$ and~$v_m$ inside~$W$, with the polygon oriented counterclockwise, we place the remaining vertices~$v_1, \dots, v_{m-1}$ such that the path~$\langle c, v_1, \dots, v_m \rangle$ utilizes wedges in~$\W$. Consequently, all vertices of~$f$ lie in~$W$ and form a strictly convex polygon.
		Every subpath~$\langle v_i, \dots, v_j \rangle$ of this path utilizes wedges in~$\W$ as well.
	\end{proof}

	\begin{restatable}[\restateref{lem:face-2-wedges}]{lemma}{faceTwoWedges}
			\label{lem:face-2-wedges}
			Let~$\W, \W' \subseteq \WL$ be disjoint consecutive wedge sets with~$|\W \cup \W'| \le |\LL|$ such that~$\W \cup \W'$ is consecutive, and let~$f$ be a bounded face of~$G$ with parent vertex~$c$ and a unique child vertex~$c'$. Then~$f$ admits a convex drawing with~$c$ at a given position such that
			\begin{enumerate*}[label=(\roman*)]
				\item for every~$v \in f \setminus \{c\}$, a path from~$c$ to~$v$ is \LL-monotone and utilizes wedges in~$\W \cup \W'$;
				\item for every~$u \in f \setminus \{c'\}$, a path from~$u$ to~$c'$ is \LL-monotone and utilizes wedges in~$\W \cup \W'$;
				\item every pair of vertices in~$f$ is connected by an~$\LL$-monotone path in~$f$.
			\end{enumerate*}
	\end{restatable}
	
	\begin{proof}[sketch]
		Let~$P$ and~$Q = \langle c, \ldots, u_q, c' \rangle$ be the two paths forming the boundary of~$f$. Refer to \cref{fig:face_monotonicity_2}. 
		We place $c$  at its given position,~$u_q$ and~$c'$ such that
		$u_q$ lies in~$W(c)$ and~$c'$ lies in the intersection~$W'(c) \cap W'(u_q)$, then invoke \cref{lem:convex_path} to embed~$P$ in~$\W'$ and the subpath~$\langle c, \ldots, u_q \rangle$ of~$Q$ in~$\W$. Any path between~$v \in P$ and~$u \in Q$ routed through~$c$ is~$l$-monotone since its subpaths lie on opposite sides of the separating line~$\overline{l}(c)$.
	\end{proof}
	
	If~$|\LL| \ge 2$, a face with one child vertex can even be drawn within a single wedge.
	
	\begin{lemma}
		\label{lem:face-one-wedge}
		Let~$|\LL| \ge 2$, let~$W \in \WL$, and let~$f$ be a bounded face of~$G$ with parent vertex~$c$ and a unique child vertex~$c'$. Then~$f$ admits a convex drawing with all vertices other than~$c$ in the interior of~$W(c)$ such that
		\begin{enumerate*}[label=(\roman*)]
			\item every vertex of~$f$ is reachable from~$c$ within~$W$;
			\item for every vertex~$u$ of~$f$, $c'$ is weakly reachable from $u$ within $W$;
			\item every pair of vertices of~$f$ is connected by an \LL-monotone path in~$f$.
		\end{enumerate*}
	\end{lemma}
	
\begin{figure}[tb]
		\captionsetup[subfigure]{justification=centering}
		\begin{subfigure}{.43\linewidth}
			\centering
			\includegraphics[page=1]{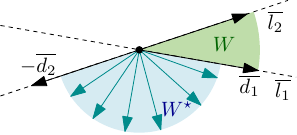}
			\subcaption{}
			\label{fig:face_one_wedge_1}
		\end{subfigure}
		\hfill
		\begin{subfigure}{.53\linewidth}
			\centering
			\includegraphics[page=2]{face_one_wedge}
			\subcaption{}
			\label{fig:face_one_wedge_2}
		\end{subfigure}
		\caption{(a)~The configuration of~\Cref{lem:face-one-wedge}. (b)~The drawing of a face~$f$ with parent~$c$ and child~$c'$ inside~$W(c)$.}
		\label{fig:face_one_wedge}
	\end{figure}

	\begin{proof}
		Let~$\overline{l_1}(c)$ and~$\overline{l_2}(c)$ be the lines of~$\overline{\LL}(c)$ that delimit~$W(c)$ in counterclockwise order, let~$\overline{d_1}$ and~$\overline{d_2}$ be their directions pointing from~$c$ along the boundary of~$W(c)$, and let~$d_1, d_2 \in \D(W)$ be the directions of~$l_1$ and~$l_2$.
		Let~$P = \langle c, u_1, \dots, u_p, c' \rangle$ and~$Q = \langle c, v_1, \dots, v_q, c' \rangle$ be the two sides of~$f$, where, as in the proof of \cref{lem:face-2-wedges}, we can choose which of them is~$P$ such that the drawing respects~$\mathcal E$, and let~$R$ be the path from~$u_1$ through~$c'$ to~$v_1$.
		If~$p = 0$, we let~$u_1 = c'$, and if~$q = 0$, we let~$v_1 = c'$; since~$f$ has at least three vertices, at most one of these cases occurs.

		Let~$W^\star$ be the open wedge from~$\overline{d_1}$ clockwise to~$-\overline{d_2}$; see \cref{fig:face_one_wedge_1}. A vector in~$W^\star$ points to the side of~$\overline{l_1}(c)$ not containing~$W$ and to the side of~$\overline{l_2}(c)$ containing~$W$, so it has a negative projection onto~$d_1$ and a positive projection onto~$d_2$.
		We place~$c'$ anywhere in the interior of~$W(c)$ and draw the edges of~$R$, from~$u_1$ to~$v_1$, parallel to~$p+q$ distinct rays in~$W^\star$ in clockwise order, starting from~$c'$ and with edges short enough that~$u_1$ and~$v_1$ lie in the interior of~$W(c)$; see \cref{fig:face_one_wedge_2}.
		Then~$f$ is a convex polygon whose vertices lie in the interior of~$W(c)$, and~$R$ is $d_2$-monotone from~$u_1$ to~$v_1$ and $d_1$-monotone from~$v_1$ to~$u_1$.

		(i)~For~$d \in \D(W)$,~$c$ is the unique lowest vertex of~$f$ in direction~$d$. As~$f$ is convex, the projection onto~$d$ increases as we walk from~$c$ along either side of~$f$ up to the highest point.
		(ii)~Since~$(c, u_1)$ points into~$W$, every vertex of~$P$ reaches~$c'$ along~$P$ in direction~$d_2$. Symmetrically, every vertex of~$Q$ reaches~$c'$ along~$Q$ in direction~$d_1$.
		(iii)~Two vertices of~$R$ are connected along~$R$, and pairs containing~$c$ by~(i).
	\end{proof}
	
	We now combine these lemmas to draw maximal chains. A chain that ends in a leaf of~$T_G$ can even be drawn within a single wedge.

	\begin{restatable}[\restateref{lem:chain}]{lemma}{chain}
		\label{lem:chain}
		Let~$M$ be a maximal chain of~$G$ with root~$c_0$, and let~$\W$ be a set of at most~$|\LL|$ consecutive wedges of~\WL. Assume that~$|\W| \ge 2$, or that~$|\W| = 1$, $|\LL| \ge 2$, and the terminal node of~$M$ is a leaf of~$T_G$. Then~$M$ admits a drawing inside~$\W(c_0)$ such that
		\begin{enumerate*}[label=(\roman*)]
			\item every vertex of~$M$ is reachable from~$c_0$ within~$\W$;
			\item if~$M$ ends in a cut vertex~$c_t$, then~$c_t$ is reachable from every vertex of~$M$ within~$\W$;
			\item every pair of vertices of~$M$ is connected by an \LL-monotone path in~$M$; and
			\item the drawing of~$M$ is planar and its bounded faces are convex.
		\end{enumerate*}
	\end{restatable}

	\begin{proof}[sketch]
		We draw the blocks of~$M$ one after the other, starting at~$c_0$.
		If~$|\W| \ge 2$, let~$W$ and~$W'$ be two consecutive wedges of~$\W$. Edges are drawn with unit length along the bisector of~$W \cup W'$, faces with one child vertex by \cref{lem:face-2-wedges}, and a face with no child vertex by \cref{lem:face-1-wedge}. Every vertex is connected to the parent vertex and to the child vertex of its block by paths that utilize wedges in~$\{W, W'\}$. Concatenating such paths yields~(i)--(iii), since~$\D(\W) \subseteq \D(\{W,W'\})$.
		If~$|\W| = 1$, say~$\W = \{W\}$, edges are drawn along the bisector of~$W$, faces with one child vertex by \cref{lem:face-one-wedge}, and a face with no child vertex by \cref{lem:face-1-wedge}. For two vertices~$u$ and~$v$ in different blocks, with~$u$ closer to~$c_0$, \cref{lem:face-one-wedge}(ii) yields a direction~$d \in \D(W)$ and a $d$-monotone path from~$u$ to the child vertex of its block, from which~$v$ is reached by a $d$-monotone path.
		In both cases, the blocks after a cut vertex~$x$ of~$M$ lie in a region with apex~$x$ that meets the earlier blocks only in~$x$, so the drawing is planar.
	\end{proof}
	
	Before addressing the high-degree faces of~$G$, we establish the following constructive lemma.
	\begin{restatable}[\restateref{lem:convex_polygon}]{lemma}{convexPolygon}
			\label{lem:convex_polygon}
			Let~$\vec{\sigma}_0, \dots, \vec{\sigma}_m$ be a sequence of
			pairwise distinct unit vectors, sorted in counterclockwise order around the
			origin.  If the smallest wedge that contains these vectors
			spans an angle strictly greater than~$\pi$,
			then there exists a convex
			polygon~$\langle u_0, \dots, u_m \rangle$, oriented counterclockwise, such that, for
			every~$i \in \{0,\dots,m\}$, the directed
			edge~$(u_i, u_{i+1})$ (with indices modulo~$m+1$) is
			parallel to and matches the orientation of~$\vec{\sigma}_i$.
	\end{restatable}
	
	We now turn our attention to the high-degree faces of~$G$. 
	
	\begin{restatable}[\restateref{lem:high_degree_faces}]{lemma}{highDegreeFaces}
			\label{lem:high_degree_faces}
			Let~$|\LL| \ge 3$, and let~$f$ be a high-degree face of~$G$ with parent vertex~$c_0$ and child vertices~$c_1, \dots, c_m$ in counterclockwise order.
			Let~$\W_{c_0}$, $\W_{c_1}$, \dots, $\W_{c_m}$ be non-empty sets of consecutive wedges that partition~\WL and appear in this counterclockwise order, where~$|\W_{c_r}| \le |\LL|$ for~$1 \le r \le m$, and let~$\mathcal{A} = \WL \setminus \W_{c_0}$.
			Let~$G' = f \cup G_{c_1} \cup \dots \cup G_{c_m}$.
                        If~$|\W_{c_0}| \le |\LL|$, then~$G'$ may also contain a subgraph~$G_{c_0}$ that meets~$f$ only in~$c_0$.
			Assume that every such subgraph~$G_c$ is drawn planarly inside~$W_c(c)$, where~$W_c = \bigcup \W_c$, that every vertex of~$G_c$ is reachable from~$c$ within~$\W_c$, and that every pair of vertices of~$G_c$ is connected by an \LL-monotone path.
			Then, $f$ admits a convex drawing such that
			\begin{enumerate}[label=(\roman*),leftmargin=*]
				\item if~$|\mathcal{A}| \le |\LL|$ and~$G'$ contains no subgraph~$G_{c_0}$, then every vertex of~$G'$ is reachable from~$c_0$ within~$\mathcal{A}$;
				\item every pair of vertices of~$G'$ is connected by an \LL-monotone path; and
				\item the drawing of~$G'$ is planar, and, if~$G'$ contains no subgraph~$G_{c_0}$, it lies inside~$\mathcal{A}(c_0)$.
			\end{enumerate}
	\end{restatable}

\begin{figure}[tb]
		\begin{minipage}[b]{.63\linewidth}
			\centering
			\includegraphics[page=1]{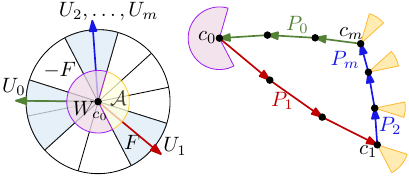}
			\caption{The construction of \cref{lem:high_degree_faces} for~$|\W_{c_0}| \ge |\LL|$.}
			\label{fig:high_degree_faces_1}
		\end{minipage}
		\hfill
		\begin{minipage}[b]{.345\linewidth}
			\centering
			\includegraphics[page=3]{high_degree_faces}
			\caption{The construction of a cactus graph~$G$ whose block-cut tree~$T_G$ has three leaves.}
			\label{fig:high_degree_face_3}
		\end{minipage}
	\end{figure}

	\begin{proof}[sketch]
		Let~$k = |\LL|$. We choose three wedges that are~$2$, $k-1$, and~$k-1$ steps apart, so that no two of them are opposite. All edges of the boundary path of~$f$ from~$c_{r-1}$ to~$c_r$ point into the same one of these wedges, which we choose at most~$k-1$ steps after the first wedge of~$\W_{c_r}$. 
		By \cref{lem:convex_polygon}, this yields a strictly convex drawing of~$f$ in which the region~$W_{c_r}(c_r)$ meets~$f$ only in~$c_r$. Since we use only three wedges, $f$ is drawn like a triangle; if~$|\W_{c_0}| \ge k$, its corners are~$c_0$, $c_1$, and~$c_m$; see \cref{fig:high_degree_faces_1}. We attach scaled-down copies of the drawings of the subgraphs~$G_{c_r}$. \arxivlncs{In the appendix}{In the full version~\cite{katsanouW-arxiv26}}, we show that conditions~(i)--(iii) hold.
	\end{proof}

	The following lemma shows that two lines suffice if~$\ell(G) = 3$, even if~$G$ has a high-degree face.

	\begin{restatable}[\restateref{lem:three-leaves}]{lemma}{threeLeaves}
		\label{lem:three-leaves}
		Let~$G$ be a cactus with~$\ell(G) = 3$ that contains a high-degree face~$f$, and let~$\mathcal E$ be an outerplanar embedding of~$G$. For every set~\LL of two lines through the origin, $G$ admits a planar \LL-monotone drawing with embedding~$\mathcal E$ whose bounded faces are convex.
	\end{restatable}

	\begin{proof}[sketch]
		The face~$f$ is the only high-degree face of~$G$, and it has exactly three cut vertices. After an affine transformation that maps the two lines to the axes, we draw the boundary of~$f$ except for one edge as an x-monotone convex arc and the three chains that hang off~$f$ inside three different quadrants; see~\cref{fig:high_degree_face_3}.
	\end{proof}

	We are now ready to show \cref{thm:upperbound}.
		
	\begin{restatable}[\restateref{thm:upperbound}]{theorem}{upperBound}
		\label{thm:upperbound}
		Let~$G$ be a connected outerplanar graph, let~$\mathcal E$ be an
		outerplanar embedding of~$G$, let~$T_G$ be the block-cut
		tree of~$G$, and let~$\ell(G)$ be the number of leaves
		of~$T_G$.  For every set~\LL of at least~$\max\{2,\lceil \ell(G)/2 \rceil\}$ lines through the origin, and of at least three lines if~$\ell(G) = 4$ and some block of~$G$ contains at least three cut vertices,
		graph~$G$ admits a planar $\LL$-monotone drawing with
		embedding~$\mathcal E$ whose bounded faces are convex.
	\end{restatable}
        
	\begin{proof}[sketch]
		Deleting all chords of~$G$ yields a spanning cactus with the same block-cut tree, in which every block has the same cut vertices as in~$G$. We draw this cactus and reinsert the chords into its convex faces, so we may assume that~$G$ is a cactus.
		If~$G$ has a high-degree face and~$|\LL| = 2$, then~$\ell(G) = 3$, and we apply \cref{lem:three-leaves}. Hence, let~$k = |\LL|$, where~$k \ge 3$ if~$G$ has a high-degree face.

		Let~$v^*$ be a leaf centroid of~$T_G$ if~$\ell(G) \ge 3$, and an inner node of the path~$T_G$ if~$\ell(G) = 2$. Then~$v^*$ is not a leaf, and every component of~$T_G - v^*$ contains at most~$\ell(G)/2 \le k$ leaves. 
		If~$v^*$ corresponds to a cut vertex, let~$c^\star$ be this cut vertex; otherwise, let~$c^\star$ be a cut vertex of the block of~$v^*$. We root~$T_G$ at the node~$v_{c^\star}$ of~$c^\star$ and, as in the proof of \cref{thm:trees}, assign to the leaves of~$T_G$ in DFS order nonempty sets of consecutive wedges that partition~\WL, such that the leaves of every component of~$T_G - v^*$ receive at most~$k$ wedges in total. Then, for every node~$w \notin \{v_{c^\star}, v^*\}$, the set~$\mathcal{A}_w$ of the wedges assigned to the leaves below~$w$ consists of at most~$k$ consecutive wedges.

		We draw~$G$ bottom-up. For every cut vertex~$c$ and every component~$G'$ of~$G_c - c$ with wedge set~$\mathcal{A}'$, we maintain that (I1)~every vertex of~$G'$ is reachable from~$c$ within~$\mathcal{A}'$, (I2)~any two vertices of~$G_c$ are connected by an \LL-monotone path, and (I3)~$G' - c$ is drawn planar inside~$\mathcal{A}'(c)$ with convex bounded faces. As the wedge sets of different components are disjoint, (I3) implies that the drawing of~$G_c$ is planar.

		A maximal chain~$M$ with root~$c_0$ and terminal node~$v_m$ is drawn by \cref{lem:chain} inside~$\mathcal{A}_{v_m}(c_0)$, which is possible since~$v_m \notin \{v_{c^\star}, v^*\}$; if~$v_m$ corresponds to a cut vertex~$c_m$, we translate the drawing of~$G_{c_m}$ such that its apex lies at~$c_m$. Every vertex of~$M$ reaches~$c_m$ within~$\mathcal{A}_{v_m}$, and~$c_m$ reaches every vertex of~$G_{c_m}$ within~$\mathcal{A}_{v_m}$, so the invariants carry over from~$c_m$ to~$c_0$.

		For a high-degree face~$f$ with parent~$c_0$ and children~$c_1$, \dots, $c_m$, we apply \cref{lem:high_degree_faces} with~$\W_{c_r} = \mathcal{A}_{v_{c_r}}$ for~$r \ge 1$ and~$\W_{c_0} = \WL \setminus \mathcal{A}_{v_f}$. If~$f \ne v^*$, then~$|\mathcal{A}_{v_f}| \le k$, and the lemma yields the invariants for the component that contains~$f$. If~$f = v^*$, then~$c_0 = c^\star$, the other components at~$c^\star$ receive at most~$k$ wedges in total, and the lemma, applied with these components as~$G_{c_0}$, yields a planar \LL-monotone drawing of~$G$.

		Finally, let~$u$ and~$w$ be vertices of different components of~$G_c - c$. By~(I1), they are reachable from~$c$ within the disjoint wedge sets of their components, so \cref{lem:reach_disjoint} yields an \LL-monotone path between them.
	\end{proof}
	
	Having established \cref{thm:upperbound}, we now prove that its bound is worst-case optimal for every~$\ell(G) \ge 2$, including the exception for~$\ell(G) = 4$; our lower-bound graphs are cacti.
	
	\begin{restatable}[\restateref{thm:lowerbound}]{theorem}{lowerBound}
			\label{thm:lowerbound}
			\begin{enumerate*}[label=(\roman*)]
				\item For every~$\ell \ge 2$, there exists an outerplanar graph~$G_\ell$ with~$\ell(G_\ell)=\ell$ in which every block contains at most two cut vertices such that, for every \LL-monotone drawing of~$G_\ell$, it holds that~$|\LL| \ge \max\{2,\lceil \ell/2 \rceil\}$.
				\item There exists an outerplanar graph~$G$ with~$\ell(G) = 4$ such that, for every planar \LL-monotone drawing of~$G$, it holds that~$|\LL| \ge 3$.
			\end{enumerate*}
	\end{restatable}
		
	\begin{proof}[sketch]
		(i)~For~$\ell \ge 3$, let~$G_\ell$ be a star with~$\ell$ leaves. Then~$\ell(G_\ell)=\ell$, and~$|\LL| \ge \lceil \ell/2 \rceil$ by \cref{thm:trees}.
		For~$\ell = 2$, the graph that consists of a $4$-cycle~$\langle s,a,t,b \rangle$ and two pendant edges~$(s,u)$ and~$(t,v)$ does not admit an \LL-monotone drawing with $|\LL|=1$.
		For~(ii), let~$G$ consist of a cycle with twelve
                vertices and pendant edges at every third vertex. In
                an \LL-monotone drawing with $|\LL|=2$ (we may assume
                $\LL\ = \{$\,x-axis, y-axis\,$\}$), the four pendant
                edges point into the four quadrants.  An analysis
                of the extremal vertices (bottom, right, top, left)
                of the cycle and the paths between them leads to a
                contradiction.
	\end{proof}

	Since every tree is a cactus, \cref{thm:upperbound} also implies the upper bound of \cref{thm:trees} for trees with at least three leaves; the direct construction in \cref{sec:trees} is simpler and serves as a blueprint for this section.
	
	\section{Conclusion and Open Problems}\label{sec:conclusion}
	
	In this paper, we showed that every tree with~$\ell$ leaves admits an $\lceil \ell/2 \rceil$-monotone drawing, which is tight, and that every outerplanar graph~$G$ admits a $\max\{2,\allowbreak\lceil \ell(G)/2 \rceil\}$-mono\-tone drawing, except if~$\ell(G) = 4$ and some block of~$G$ contains at least three cut vertices, in which case it admits a $3$-monotone drawing; these bounds are tight.
	Both algorithms produce planar drawings that respect the given embedding. We neither optimize the area nor the angular resolution of the drawings; investigating the trade-offs between the number of lines, the angular resolution, and the required grid area is an interesting direction for future work.
	While all $3$-connected planar graphs are known to be $3$-monotone, whether they are all $2$-monotone remains open. For planar graphs, the required number of lines is unknown.
	
	\bibliographystyle{splncs04}
	\bibliography{bibliography}
	
	\clearpage
\arxivlncs{}{\end{document}}

	\appendix
	
	\section*{Appendix}
	
\begin{figure}[tb]
		\centering
		\includegraphics[page=1]{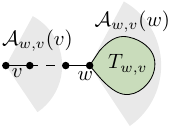}
		\caption{Drawing the path~$P=\langle v,\dots,w\rangle$ from the apex~$v$ in the proof of \cref{thm:trees}.}
		\label{fig:tree_branch_regions_1}
	\end{figure}

	\trees*
	\label{thm:trees*}
	\begin{proof}
		Regarding the second statement of the theorem,
		$|\LL| \geq \ell'$ is implied
        by \cref{lem:max-degree-2k}.
        We now show that this number is also sufficient, using
        \cref{thm:trees_characterization}. 
        If~$T$ has no branching vertex, then~$T$ is a path or a single vertex, and drawing it on a line parallel to a line of~\LL yields an \LL-monotone drawing. Hence, we assume that~$T$ has a branching vertex, so~$\ell \ge 3$.
        Let~$\LL=\{l_0,\dots,l_{\ell'-1}\}$ and 
        let~$\WL$ denote the set of~$2\ell'$ wedges induced by the lines in~$\overline{\LL}$.
        \cref{fig:tree_branch_regions_3} shows an \LL-monotone drawing of a tree with eight leaves.
        
        We associate the~$\ell$ leaves of~$T$ with the wedges of~$\WL$ as
        follows.
        Let~$u$ be a leaf centroid of~$T$. Root~$T$ at~$u$ and perform a depth-first search (DFS) traversal.
        Enumerate the leaves in the order in which they are encountered by the
        DFS traversal and denote them by~$\lambda_0, \lambda_1,\dots,\lambda_{\ell-1}$.
        Assign to each leaf~$\lambda_i$ the wedge~$W_i\in \WL$; that is, the
        leaf path terminating at~$\lambda_i$ will utilize wedge~$W_i$.
        If~$\ell$ is odd, assign the remaining wedge~$W_{2\ell'-1}$ to the leaf~$\lambda_{\ell-1}$.
        The DFS visits the children of every vertex in the counterclockwise order given by the embedding of~$T$, starting, at every vertex other than the root, with the child that follows the parent in this order, and we number the wedges~$W_0, \dots, W_{2\ell'-1}$ of~$\WL$ in counterclockwise order. Hence, the leaves of every subtree receive consecutive wedges, and the subtrees of every vertex receive their wedges in the cyclic order of the embedding, so the drawing respects the embedding of~$T$.
        Let~$\mathcal{A}_{\lambda_i}$ denote the set of one or two wedges assigned to leaf~$\lambda_i$. 
        
        We now describe the construction of the drawing.
        Fix the root vertex~$u$ at the origin and traverse~$T$ once more using a
        DFS traversal.
        Let~$P=\langle v,\dots,w\rangle$ be a branch or a leaf path of~$T$ such
        that~$v$ is visited before~$w$ in the DFS traversal, and assume that~$v$
        has already been placed.
        We draw the remaining vertices of~$P$ as follows.
        Let~$\mathcal{A}_{w,v}$ be the union of the wedge sets assigned to the leaves of the
        subtree~$T_{w,v}$.
        If~$w$ is a leaf, then~$\mathcal{A}_{w,v}$ coincides with~$\mathcal{A}_w$.
		Since~$v$ is visited before~$w$, the subtree~$T_{w,v}$ lies in a component of~$T - u$ and, hence, contains at most~$\ell/2$ leaves; if~$\ell$ is odd, these are at most~$\ell'-1$ leaves, one of which may be assigned two wedges. Thus,~$|\mathcal{A}_{w,v}| \le \ell'$.
Refer to \cref{fig:tree_branch_regions_1}. We draw path~$P$ inside the region~$\mathcal{A}_{w,v}(v)$, using unit-length edges whose directions coincide with the bisector of~$\mathcal{A}_{w,v}$, rotated slightly if this bisector lies on a line of~$\overline{\LL}$.
        
        Observe that the oriented
         path~$P=\langle v,\dots,w\rangle$ utilizes only wedges in~$\mathcal{A}_{w,v}$; that is,~$\W_P \subseteq \mathcal{A}_{w,v}$.
        Since~$\mathcal{A}_{w,v}\cup \mathcal{A}_{v,w}=\WL$ and~$\mathcal{A}_{w,v}\cap \mathcal{A}_{v,w}=\emptyset$, the
        reverse path~$\langle w,\dots,v\rangle$, which utilizes the wedges opposite
        to those in~$\W_P$, satisfies~$\W_P^- \subseteq \mathcal{A}_{v,w}$.
        Consequently,~$\W_P \cap \mathcal{A}_{v,w}=\emptyset$ and~$\W_P^- \cap \mathcal{A}_{w,v}=\emptyset$.
        Moreover, every branch or leaf path in~$T_{v,w}$, oriented away from~$v$, utilizes only wedges of~$\mathcal{A}_{v,w}$. If it is oriented away from the root, it utilizes wedges assigned to leaves of the subtree below its last vertex; otherwise, it utilizes the wedges opposite to those assigned to the leaves below its first vertex, which are assigned to leaves outside this subtree. In both cases, these leaves lie in~$T_{v,w}$.
        Therefore,~$\W_{v,w} \subseteq \mathcal{A}_{v,w}$ and, similarly,~$\W_{w,v} \subseteq \mathcal{A}_{w,v}$.
       
        By construction, every branch and every leaf path of~$T$ is~$\LL$-monotone; hence, the drawing satisfies condition~(a) of
        \cref{thm:trees_characterization}.
        Moreover, condition~(b) is satisfied because distinct wedge sets are
        assigned to the leaf paths of~$T$, and each leaf path utilizes exactly its
        assigned wedge set.
        Finally, for condition~(c), we have shown that for any branch or leaf path~$P=\langle v,\dots,w\rangle$, it holds that~$\W_P \cap \W_{v,w}=\emptyset$ and~$\W_P^- \cap \W_{w,v}=\emptyset$.
        Recall that~$R_{v,w}$~$= \W_P(v) \cap \W_P^-(w)$.
        Since the wedge sets~$\W_P$ and~$\W_{v,w}$ are disjoint, the interiors of the regions~$\W_{v,w}(v)$ and~$\W_P(v) \supseteq R_{v,w}$ are disjoint; similarly, the interiors of~$\W_{w,v}(w)$ and~$\W_P^-(w) \supseteq R_{v,w}$ are disjoint.
        Thus, the drawing also satisfies condition~(c).
	\end{proof}

	\wedgesDisjoint*
	\label{lem:wedges_disjoint*}
	\begin{proof}
		Since~$P$ and~$Q$ are~$\LL$-monotone, it follows that~$|\W_{P}| \le |\LL|$ and~$|\W_{Q}| \le |\LL|$. 
		As these two wedge sets are disjoint, there exists a line~$\overline{l} \in \overline{\LL}$ such that~$\W_{P}$ and~$\W_{Q}$ lie in opposite open half-planes defined by~$\overline{l}$. 
		Indeed, the wedges not in~$\W_P$ are at least~$|\LL|$ consecutive wedges that contain~$\W_Q$, so~$|\LL|$ consecutive ones among them contain~$\W_Q$. These form an open half-plane delimited by a line~$\overline{l} \in \overline{\LL}$, and~$\W_P$ lies in the opposite one.
		Let~$P^-$ denote the reversal of the path~$P$, routing from~$u$ to~$x$. 
		The edges of~$P^-$ utilize the wedge set~$\W_{P}^-$, which consists of the wedges directly opposite to those in~$\W_{P}$. 
		Consequently, every edge of the full path~$\Pi$ utilizes a wedge of~$\W_{P}^- \cup \W_{Q}$. 
		Since~$\W_{P}$ and~$\W_{Q}$ lie in opposite half-planes with respect to~$\overline{l}$, taking the opposite wedges of~$\W_{P}$ reflects them across the origin, meaning that~$\W_{P}^-$ and~$\W_{Q}$ now lie within the same open half-plane delimited by~$\overline{l}$. 
		The~$|\LL|$ wedges in this half-plane are consecutive, so the smallest set~$\W_\Pi$ of consecutive wedges that contains~$\W_{P}^- \cup \W_{Q}$ lies in this half-plane, too.
		Therefore,~$\overline{l}$ does not intersect the interior of~$\W_\Pi$, which, by \cref{lem:monot_external_dir}, guarantees that the path~$\Pi$ is~$\LL$-monotone.
	\end{proof}
	
	\reachDisjoint*
	\label{lem:reach_disjoint*}
	\begin{proof}
		As in the proof of \cref{lem:wedges_disjoint}, there is a line~$\overline{l} \in \overline{\LL}$ such that~$\W$ and~$\W'$ lie in opposite open half-planes delimited by~$\overline{l}$.
		Let~$d \in \D_\LL$ be the direction of~$l$ that points towards the half-plane containing~$\W'$. Then~$d \in \D(\W')$ and~$-d \in \D(\W)$.
		Reversing a $(-d)$-monotone path from~$x$ to~$u$ and appending a $d$-monotone path from~$x$ to~$v$ yields a $d$-monotone path from~$u$ to~$v$.
	\end{proof}

        In the following two lemmas, we let~$\|z\|$ denote the
        Euclidean length of the vector~$z$.
        
	\convexPath*
	\label{lem:convex_path*}
	\begin{proof}
          Let~$l^\star$ be the straight line passing through points~$x$ and~$y$; see \cref{fig:convex_path_1}.
          This line partitions~$W(x)$ into two smaller wedges~$W^{(1)}(x)$ and~$W^{(2)}(x)$.
          We assume, without loss of generality, that~$W^{(1)}(x)$ is the narrower of the two.
		We construct the path~$P=\langle x = p_0, p_1, \dots, p_{n},  p_{n+1} = y \rangle$ such that all of its~$n+1$ edges have identical length ($L$, defined below) and all internal angles~$\angle p_0 p_1 p_2, \dots, \angle p_{n-1} p_n p_{n+1}$ are equal ($\phi$, defined below).
		We choose the initial edge~$(x, p_1)$ to have a direction
                that lies within the narrower wedge~$W^{(1)}(x)$.
		To place the path on the other side of~$l^\star$, we instead choose the direction of~$(x, p_1)$ in~$W^{(2)}(x)$ such that its angle with~$l^\star$ is smaller than the angle of~$W^{(1)}(x)$; the argument below is the same with the roles of~$W^{(1)}(x)$ and~$W^{(2)}(x)$ exchanged.
                Now we can apply \cref{lem:path_arc} below, which for
		$\theta=\angle y x p_{1}$ yields $\phi = \pi - 2\theta/n$ and
                $L = \|x - y\| \cdot \frac{\sin(\theta/n)}{\sin(\theta(n+1)/n)}$.
		
		The polygon~$\langle x, p_1, \dots, p_{n}, y, x \rangle$ is convex by construction.
		It remains to argue that the path utilizes only wedges in~\W. 
		Since~$(x, p_1)$ is chosen strictly within~$W^{(1)}(x)$ and the path is constructed with equal angles, the direction of the final segment~$(p_{n}, y)$ is symmetric to the direction of~$(x, p_1)$ with respect to~$l^\star$; see \cref{fig:convex_path_2}.
                Hence, as~$\theta$ is smaller than the angle of~$W^{(2)}(x)$, the segment~$(p_{n}, y)$ lies in~$W^{(2)}(p_{n})$.
                The directions of the other edges lie between these two directions, and~$W$ is convex since~$|\W| \le |\LL|$.
                This construction ensures that, for every edge in~$P$, if we translate its start point to the origin, then the edge lies in~$W$.
                Thus, $P$ utilizes only wedges in~\W.
        \begin{figure}[tb]
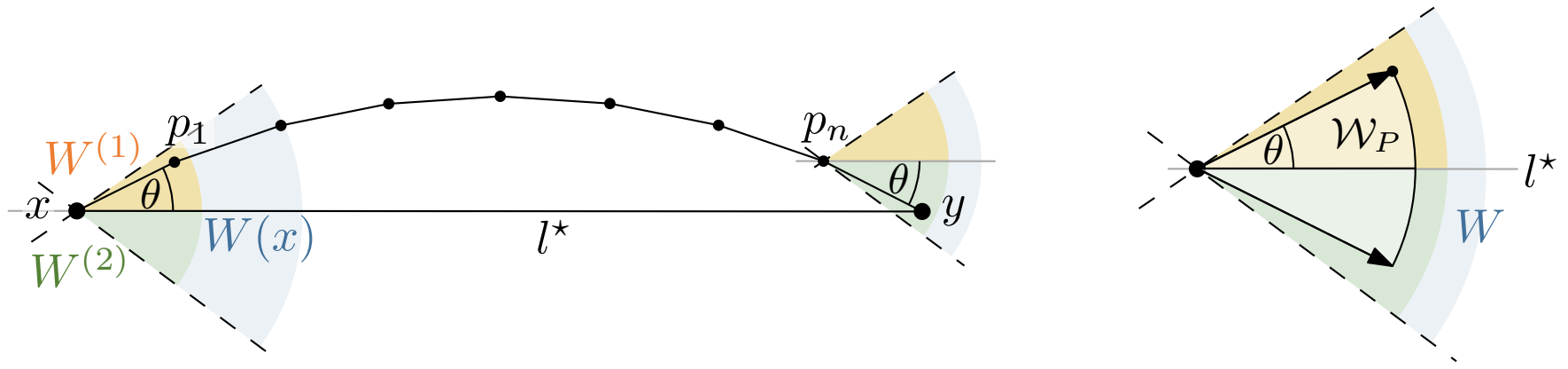

              	\captionsetup[subfigure]{justification=centering}
              	\begin{subfigure}{.61\linewidth}
              		\centering
              		\includegraphics[page=2]{path_arc}
              		\subcaption{}
              		\label{fig:convex_path_1}
              	\end{subfigure}
              	\hfill
              	\begin{subfigure}{.35\linewidth}
              		\centering
              		\includegraphics[page=3]{path_arc}
              		\subcaption{}
              		\label{fig:convex_path_2}
              	\end{subfigure}
              	\caption{(a)~The construction of path $P$ in the proof of \cref{lem:convex_path}. (b)~Path $P$ utilizes wedges in \W.}
              	\label{fi:convex_path}
        \end{figure}
	\end{proof}
	
	\begin{lemma}\label{lem:path_arc}
		Let~$P = \langle p_0, p_1, \dots, p_{n+1} \rangle$ be a path embedded such that the polygon~$\langle p_0, p_1, \dots, p_{n+1}, p_0 \rangle$ is convex, and let~$\theta = \angle p_{n+1} p_0 p_1$.
		If all~$n+1$ edges of~$P$ have identical length~$L$ and all~$n$ internal angles are equal, i.e.,
		$\phi = \angle p_0 p_1 p_2 = \dots = \angle p_{n-1} p_n p_{n+1}$, then
		\begin{enumerate}[label=(\roman*)]
			\item $\phi = \pi - \frac{2\theta}{n}$, and
			\item $L = \|p_{n+1} - p_0\| \cdot \frac{\sin(\theta/n)}{\sin\left(\frac{n+1}{n}\theta\right)}$.
		\end{enumerate}
	\end{lemma}
	
	\begin{proof}
		Refer to \cref{fig:path_arc}. Because path~$P$ is constructed with uniform edge lengths and equal internal angles, its vertices~$p_0, p_1, \dots, p_n, p_{n+1}$ lie on a circular arc. Let~$O$ and~$R$ denote the center and the radius of the underlying circle, respectively. 
		\begin{figure}[tb]
			\centering
			\includegraphics[page=1]{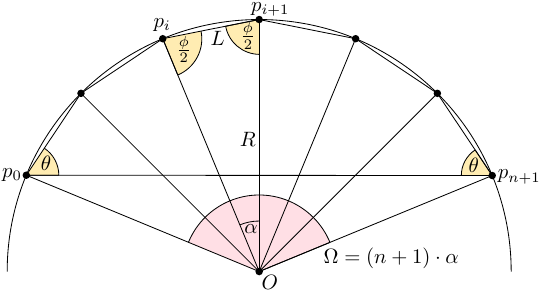}
			\caption{The construction of \cref{lem:path_arc}.}
			\label{fig:path_arc}
		\end{figure}
		To establish property~(i), consider the closed polygon formed by the path and the chord line, given by~$\langle p_0, p_1, \dots, p_n, p_{n+1} \rangle$. This polygon has~$n+2$ vertices, implying the sum of its interior angles is exactly~$\pi n$. By symmetry, the two base angles adjacent to the chord at~$p_0$ and~$p_{n+1}$ are both equal to~$\theta$, while the remaining~$n$ internal vertices share the angle~$\phi$, which immediately yields~$\phi = \pi - 2\theta/n$.
		
		To establish property~(ii), let~$\alpha = \angle p_0 O p_{1} = \dots = \angle p_n O p_{n+1}$.
		Clearly, $\alpha = \pi - \phi = 2\theta/n$.
		Letting~$\Omega = \angle p_0 O p_{n+1}$, we obtain~$\Omega = (n+1)\alpha = 2\theta(n+1)/{n}$.
		From the isosceles triangles~$\triangle p_i O p_{i+1}$,
		for~$0 \le i \le n$, we have~$L = 2R \sin(\alpha/2) = 2R \sin(\theta/n)$.
		Similarly, considering the isosceles triangle~$\triangle p_0 O p_{n+1}$ and letting~$d = \|p_{n+1} - p_0\|$ denote the chord distance, we obtain~$d = 2R \sin(\Omega/2) = 2R \sin(\theta(n+1)/n)$.
		Eliminating~$2R$ yields the desired uniform edge
		length~$L = \|p_{n+1} - p_0\| \cdot \frac{\sin(\theta/n)}{\sin(\theta(n+1)/n)}.$
		This completes the proof.
	\end{proof}
	
	\faceTwoWedges*
	\label{lem:face-2-wedges*}
	\begin{proof}
		Let~$W=\bigcup \W$ and~$W'=\bigcup \W'$ be the smallest wedges that contain the wedges in~$\W$ and~$\W'$, respectively, and let~$\overline{l} \in \overline{\LL}$ be the line delimiting~$W$ and~$W'$. Refer to \cref{fig:face_monotonicity_2}.
		Let~$P = \langle c, v_1, \dots, v_p, c' \rangle$ and~$Q = \langle c, u_1, \dots, u_q, c' \rangle$ be the two paths forming the boundary of~$f$; since the construction below works for all~$p$ and~$q$, we can choose which of the two paths is~$Q$ such that the drawing respects~$\mathcal E$.
		We start by placing the parent vertex~$c$ at its given point in the plane. We, then, place~$u_q$ and~$c'$ such that, if~$q \ge 1$,
		$u_q$ lies in the interior of~$W(c)$ and~$c'$ lies in the interior of the intersection~$W'(c) \cap W'(u_q)$; see \cref{fig:face_monotonicity_2}. Observe that this intersection is always non-empty. 
		
		To construct the boundary, we apply \cref{lem:convex_path} twice. First, we apply it to the full path~$P$ using the fixed endpoints~$c$ and~$c'$ within the wedge~$W'$, such that the inner vertices of~$P$ lie on the side of the line through~$c$ and~$c'$ that is disjoint from the interior of~$W(c)$. By construction, path~$P$ utilizes wedges in~$\W'$, and since~$|\W'| \le |\LL|$, it is \LL-monotone by \cref{lem:monot_external_dir}.
		Second, we apply \cref{lem:convex_path} to the subpath~$\langle c, u_1, \dots, u_q \rangle$ of~$Q$ using fixed endpoints~$c$ and~$u_q$ within~$W$, such that its inner vertices lie on the side of the line through~$c$ and~$u_q$ that does not contain~$c'$. We complete the construction by adding the edge~$(u_q, c')$.
		If~$p = 0$, then~$P$ is the edge~$(c, c')$, and if~$q = 1$, then~$\langle c, u_1 \rangle$ is the edge~$(c, u_1)$; we draw such a path as a straight-line segment instead of applying \cref{lem:convex_path}. If~$q = 0$, then~$u_q = c$ and~$Q$ is the edge~$(c, c')$. Since~$G$ has no multiple edges,~$p$ and~$q$ are not both~$0$. Because the subpath~$\langle c, \dots, u_q \rangle$ utilizes wedges in~$\W$ and the edge~$(u_q, c')$ utilizes wedges in~$\W'$, the entire path~$Q$ utilizes wedges in~$\W \cup \W'$ and is \LL-monotone by \cref{lem:monot_external_dir}.
		Since~$P$ and~$\langle c, \dots, u_q \rangle$ are drawn convex and bend away from each other, $f$ is drawn as a convex polygon; in particular, the drawing is planar.
		
		Since both boundary paths~$P$ and~$Q$ utilize wedges in~$\W \cup \W'$, conditions~(i) and~(ii) are satisfied, as the paths in question are subpaths of~$P$ or~$Q$. Furthermore, for any vertices~$v \in P$ and~$u \in Q$  with~$u \ne c'$, the path between them passing through~$c$ is~\LL-monotone by \cref{lem:wedges_disjoint}, since the subpaths~$\langle c, \dots, v \rangle$ and~$\langle c, \dots, u \rangle$ are \LL-monotone and utilize wedges of the disjoint sets~$\W'$ and~$\W$, respectively; for~$u = c'$, both vertices lie on~$P$.
		Two vertices on the same boundary path are connected by a subpath of~$P$ or of~$Q$, which utilizes wedges in~$\W'$ or in~$\W \cup \W'$, respectively, and is therefore \LL-monotone by \cref{lem:monot_external_dir}.
		Finally, by~(i) and~(ii), and since the region~$(W \cup W')(c)$ is convex as~$|\W \cup \W'| \le |\LL|$, the face~$f$ lies inside~$(W \cup W')(c)$ and inside the reflection of~$(W \cup W')(c')$ at~$c'$.

	\end{proof}
	
	\chain*
	\label{lem:chain*}
	\begin{proof}
		Let~$c_0, x_1, \dots, x_t$ be the cut vertices of~$M$ in the order of~$M$, and let~$x_0 = c_0$. For~$s = 1, \dots, t$, the $s$-th block of~$M$ has parent vertex~$x_{s-1}$ and child vertex~$x_s$. If the terminal node of~$M$ is a leaf of~$T_G$, the last block contains~$x_t$ as its only cut vertex; otherwise,~$M$ ends in~$c_t = x_t$.
		We draw the blocks one after the other, starting at~$c_0$, which we place at an arbitrary point in the plane.
		
		Case~$|\W| \ge 2$. Let~$W$ and~$W'$ be two consecutive wedges of~$\W$.
		An edge with parent vertex~$x_{s-1}$ is drawn with unit length along the angular bisector of~$W(x_{s-1}) \cup W'(x_{s-1})$, rotated slightly if this bisector lies on a line of~$\overline{\LL}$; it utilizes a wedge in~$\{W, W'\}$, since it points into the interior of~$W(x_{s-1}) \cup W'(x_{s-1})$.
		A face with one child vertex is drawn by \cref{lem:face-2-wedges} with the wedge sets~$\{W\}$ and~$\{W'\}$, and a face with no child vertex by \cref{lem:face-1-wedge} with~$\{W\}$.
		Hence, in every block, the path from its parent vertex to any of its vertices and, if the block has a child vertex, the path from any of its vertices to its child vertex are \LL-monotone and utilize wedges in~$\{W, W'\}$.
		For a vertex~$v$ of~$M$, concatenating these paths yields a path from~$c_0$ to~$v$ and, if~$M$ ends in~$c_t$, a path from~$v$ to~$c_t$ that utilize wedges in~$\{W, W'\}$. Such a path is $d$-monotone for every~$d \in \D(\{W,W'\}) \supseteq \D(\W)$, which shows~(i) and~(ii).
		For~(iii), two vertices in the same block are connected by \cref{lem:face-2-wedges,lem:face-1-wedge}. For two vertices~$u$ and~$v$ in different blocks, with~$u$ closer to~$c_0$, the path from~$u$ to the child vertex of its block, followed by the path from there to~$v$, utilizes wedges in~$\{W, W'\}$ and, hence, is \LL-monotone.
		For planarity, all edges of~$M$, oriented away from~$c_0$, point into the interior of~$W \cup W'$. This holds for the edges drawn along the (slightly rotated) bisector by construction, and for the edges of the faces by the construction in \cref{lem:face-2-wedges,lem:face-1-wedge}.
		Hence, by the paths above, every vertex of the $s$-th block other than~$x_s$ lies in the interior of the reflection of~$(W \cup W')(x_s)$ at~$x_s$, and every vertex of a later block other than~$x_s$ lies in the interior of~$(W \cup W')(x_s)$.
		Since both regions are convex, the $s$-th block and the later blocks meet only in~$x_s$, so the drawing is planar. Its bounded faces are convex by \cref{lem:face-2-wedges,lem:face-1-wedge}.
		
		Case~$|\W| = 1$. Then~$\W = \{W\}$, $|\LL| \ge 2$, and the terminal node of~$M$ is a leaf of~$T_G$, so~(ii) is void.
		An edge~$(x_{s-1}, y)$ is drawn with unit length along the bisector of~$W(x_{s-1})$, rotated slightly if this bisector lies on a line of~$\overline{\LL}$, a face with one child vertex by \cref{lem:face-one-wedge} with apex~$x_{s-1}$, and a face with no child vertex by \cref{lem:face-1-wedge} with~$\{W\}$ and apex~$x_t$.
		Every block lies in~$W(x_{s-1}) \cup \{x_{s-1}\}$, and it meets~$W(x_s)$ only in its child vertex~$x_s$, while~$W(x_s)$ contains all later blocks; hence, the drawing is planar. Its bounded faces are convex by \cref{lem:face-one-wedge,lem:face-1-wedge}.
		For an edge, this is clear. For a face drawn by \cref{lem:face-one-wedge}, we use the notation of its proof. If we traverse the boundary of the face such that~$R$ is traversed from~$u_1$ to~$v_1$, then the edge directions turn clockwise. Hence, since the face is convex, it lies in the cone at~$x_s$ that we sweep by rotating the direction of the edge leaving~$x_s$ in this traversal clockwise until we reach the direction opposite to the edge entering~$x_s$. The first direction lies in~$W^\star$ or, if~$x_s = v_1$, in~$-W$, and the second one lies in~$-W^\star$ or, if~$x_s = u_1$, in~$-W$. Rotating clockwise, the wedges~$W^\star$, $-W$, and~$-W^\star$ follow each other, and~$-W^\star$ ends at~$\overline{d_2}$, so the sweep does not meet~$W$. Hence, the face meets~$W(x_s)$ only in~$x_s$.
		
		Every vertex of the $s$-th block is, by construction, reachable from~$x_{s-1}$ within~$W$; for edges and the final face, this holds since the relevant paths utilize only~$W$. Concatenating, for a fixed~$d \in \D(W)$, $d$-monotone paths from~$x_0$ to~$x_1$, from~$x_1$ to~$x_2$, and so on, shows that every vertex of~$M$ is reachable from~$c_0$ within~$W$.
		Two vertices in the same block are connected by an \LL-monotone path by part~(iii) of \cref{lem:face-one-wedge} or by \cref{lem:face-1-wedge}. Let~$u$ lie in the $s$-th block and~$v$ in a later block. By \cref{lem:face-one-wedge}(ii), or trivially if the $s$-th block is an edge, there are a direction~$d \in \D(W)$ and a $d$-monotone path from~$u$ to~$x_s$. Since~$v$ is reachable from~$x_s$ within~$W$, there is a $d$-monotone path from~$x_s$ to~$v$, and the concatenation is $d$-monotone.

	\end{proof}
	
	\convexPolygon*
	\label{lem:convex_polygon*}
	\begin{proof}
		For~$i \in \{0,\dots,m\}$, let~$h_i$ be the closed halfplane
		that contains the unit disk and whose bounding line
		is tangent to the unit circle, is parallel to~$\vec{\sigma}_i$, and has the disk to its left when directed as~$\vec{\sigma}_i$.  Then the boundary~$\partial H$ of
		~$H=h_0 \cap \dots \cap h_m$ is the desired convex polygon.
		Clearly,~$\partial H$ is convex.  Since the smallest wedge
		that contains the vectors
		$\vec{\sigma}_0, \dots, \vec{\sigma}_m$ spans an angle
		strictly greater than~$\pi$,~$H$ is bounded and each tangent
		contributes to~$\partial H$ a line segment (which contains
		the tangent point).  The vertices of~$\partial H$ are
		exactly the intersection points of the pairs of consecutive
		tangents around the circle.
	\end{proof}
	
	\highDegreeFaces*
	\label{lem:high_degree_faces*}
	\begin{proof}
			Let~$k = |\LL|$.
			For two wedges~$W$ and~$W'$, let~$\mathrm{dist}(W, W')$ be the number of steps from~$W$ to~$W'$ when we walk counterclockwise from wedge to wedge; for example,~$\mathrm{dist}(W, -W) = k$.
			We call a set of consecutive wedges starting at~$W$ and ending at~$W'$ \emph{short} if~$\mathrm{dist}(W, W') \le k-1$.
			For a short set~$\mathcal{S}$, we have~$\D(\mathcal{S}) \ne \emptyset$, and, by \cref{lem:monot_external_dir}, every path that utilizes wedges of~$\mathcal{S}$ is $d$-monotone for every~$d \in \D(\mathcal{S})$.
			
			For~$0 \le r \le m$, let~$F_r$ and~$L_r$ be the first and the last wedge of~$\W_{c_r}$ (in counterclockwise order), and let~$P_r$ be the counterclockwise path along~$f$ that starts in~$c_{r-1}$ and ends in~$c_r$; indices are taken modulo~$m+1$.
			
			We number the wedges by integers in counterclockwise order, starting with~$0$ for~$F_0$; the numbers~$t$ and~$t + 2k$ denote the same wedge.
			We use the same letters for the wedges~$F_r$ and~$L_r$ and their numbers, so~$0 = F_0 \le L_0 < F_1 \le L_1 < \dots < F_m \le L_m < 2k$.
			
			We first describe how we choose the directions of the boundary paths.
			All edges of~$P_r$ will point into one wedge~$U_r$, where the wedges~$U_0, \dots, U_m$ satisfy the following conditions.
			\begin{enumerate}[label=\textbf{(D\arabic*)},leftmargin=*,ref=(D\arabic*)]
				\item \label{D:window} $\mathrm{dist}(F_r, U_r) \le k-1$ for every~$r$, that is,~$U_r$ has a number~$u_r$ with~$F_r \le u_r \le F_r + k - 1$;
				\item \label{D:gap} $\mathrm{dist}(U_r, U_{r+1}) \le k-1$ for every~$r$, and the wedges~$U_0, U_1, \dots, U_m, U_0$ go around the origin exactly once, that is, these distances sum up to~$2k$;
				\item \label{D:opp} no two of the wedges~$U_0, \dots, U_m$ are opposite.
			\end{enumerate}
			In both cases below, we only use three wedges; they are~$2$,~$k-1$, and~$k-1$ steps apart, so no two of them are opposite, which shows~\ref{D:opp}.
			
			\textbf{Main case~$|\W_{c_0}| \ge k$, that is,~$|\mathcal{A}| \le k$.}
			Let~$F = F_1$ be the first wedge of~$\mathcal{A}$.
			We let~$U_1 = F$, we let~$U_2 = \dots = U_m$ be the wedge just before~$-F$, and we let~$U_0$ be the wedge just after~$-F$; see \cref{fig:high_degree_faces_1}.
			So the path from~$c_0$ to~$c_1$ points into the first wedge of~$\mathcal{A}$, the paths from~$c_1$ to~$c_m$ all point into one wedge, and~$f$ is drawn like a triangle with corners~$c_0$,~$c_1$, and~$c_m$.
			For~\ref{D:window}, note that~$\mathcal{A}$ consists of at most~$k$ wedges starting with~$F$, and the wedge just before~$-F$ is the~$k$-th wedge starting with~$F$; hence~$\mathrm{dist}(F_r, U_r) \le k-1$ for~$1 \le r \le m$.
			Since~$m \ge 2$,~$\mathcal{A}$ has at least two wedges, so~$F_0$ (the wedge after the last wedge of~$\mathcal{A}$) is at least two and at most~$k$ steps after~$F$, and hence~$\mathrm{dist}(F_0, U_0) \le k-1$.
			For~\ref{D:gap}, the distances from~$U_1$ to~$U_2$, from~$U_m$ to~$U_0$, and from~$U_0$ to~$U_1$ are~$k-1$,~$2$, and~$k-1$, respectively, and the others are zero.
			
			\begin{figure}[tb]
				\centering
				\includegraphics[page=2]{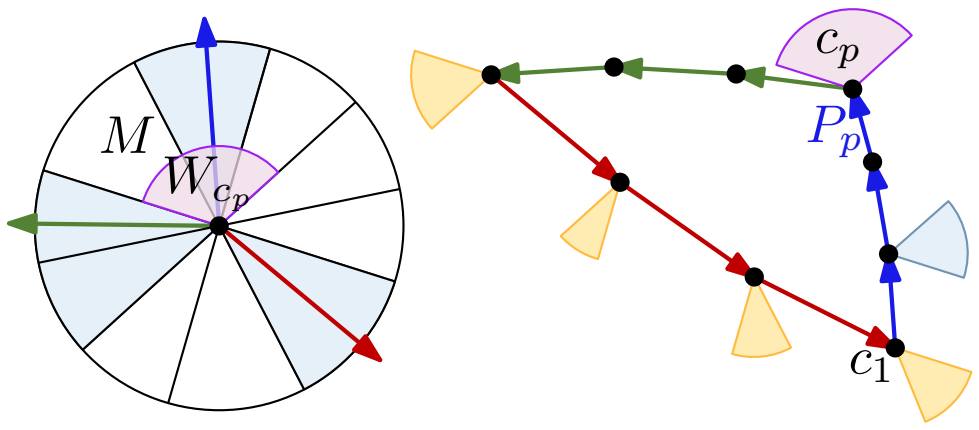}
				\caption{The construction of \cref{lem:high_degree_faces} in the root case, where all sets~$\W_{c_r}$ have at most~$k$ wedges.}
				\label{fig:high_degree_faces_2}
			\end{figure}
			
			\textbf{Root case~$|\W_{c_0}| \le k$.}
			Then all~$m+1 \ge 3$ sets~$\W_{c_r}$ have at most~$k$ wedges, and at most one of them has exactly~$k$ wedges.
			Let~$\W_{c_p}$ be this set if it exists, and an arbitrary set otherwise, and let~$M$ be the last wedge of~$\W_{c_p}$.
			Let~$T$ consist of the wedge just before~$M$, the wedge just after~$M$, and~$-M$, and, for every $0 \le r \le m$, let~$U_r$ be the first wedge of~$T$ at or after~$F_r$; see \cref{fig:high_degree_faces_2}.
			Any~$k-1$ consecutive wedges contain a wedge of~$T$, which shows~\ref{D:window}.
			The three wedges of~$T$ lie in different sets. The wedges just before and just after~$M$ do, since~$M$ is the last wedge of~$\W_{c_p}$; and a set containing~$-M$ and one of them would contain the~$k$ wedges from one to the other, so it would be a set with exactly~$k$ wedges other than~$\W_{c_p}$.
			Hence,~$U_{r+1} = U_r$ unless~$\W_{c_r}$ contains a wedge~$t$ of~$T$; then~$U_r = t$, and~$U_{r+1}$ is the next wedge of~$T$ after~$t$.
			So~$U_0, \dots, U_m, U_0$ run once through the three wedges of~$T$ in counterclockwise order, which shows~\ref{D:gap}, since consecutive wedges of~$T$ are~$2$ or~$k-1$ steps apart.
			
			We now proceed with the construction of the face.
			By~\ref{D:gap}, the paths that use the same wedge~$U$ are consecutive along~$f$.
			We choose distinct rays from the origin inside~$U$ in counterclockwise order, one for each edge of these paths, and we let the $i$-th of these edges along~$f$ point in the direction of the $i$-th ray.
			Then, as we walk along~$f$, the edge directions turn counterclockwise exactly once around the origin, and, by~\ref{D:gap}, two consecutive ones enclose an angle less than~$\pi$.
			Hence, \cref{lem:convex_polygon} yields a convex polygon with these edge directions, which is strictly convex since the directions are pairwise distinct; we use it as the drawing of~$f$.
			
			The interior angle of~$f$ at~$c_r$ lies in the wedges from~$U_{r+1}$ to~$-U_r$.
			By~\ref{D:window},~$F_r$ comes strictly after~$-U_r$, and~$L_r$ comes before~$F_{r+1}$, which is at or before~$U_{r+1}$.
			So the wedges of~$\W_{c_r}$ lie strictly between~$-U_r$ and~$U_{r+1}$, outside the interior angle, and the region~$W_{c_r}(c_r)$ meets~$f$ only in~$c_r$.
			We attach the drawing of~$G_{c_r}$ at~$c_r$, scaled down such that the subgraphs are pairwise disjoint.
			If there is no subgraph~$G_{c_0}$, then the interior angle at~$c_0$ lies in the wedges from~$U_1$ to~$-U_0$, which have the numbers from~$u_1 \ge F_1$ to~$u_0 + k \le 2k - 1$ by~\ref{D:window}; so it lies in~$\mathcal{A}$. Hence,~$f$ lies inside~$\mathcal{A}(c_0)$, and we scale the subgraphs such that they also lie inside~$\mathcal{A}(c_0)$.
			This shows~(iii); scaling preserves reachability and monotone paths.
			
			\medskip
			\noindent\textbf{Monotone paths along~$f$.}
			The boundary of~$f$ consists of three parts, one for each of the three wedges that we use; they are~$2$,~$k-1$, and~$k-1$ steps apart.
			Two vertices~$u$ and~$v$ of~$f$ are inner vertices of at most two of these parts, so one of the two paths between~$u$ and~$v$ along~$f$ avoids a whole part.
			This path uses at most two of the three wedges, which are at most~$k-1$ steps apart, so it utilizes a short set and is monotone.
			
			\medskip
			\noindent\textbf{Vertices of the subgraphs.}
			Two vertices of the same subgraph~$G_c$ are connected by an \LL-monotone path by assumption.
			For a vertex~$v$ of~$f$ that is not a cut vertex, let~$G_v$ consist of~$v$ alone.
			Now let~$x \in G_c$ and~$y \in G_{c'}$ with~$c \ne c'$, where~$c$ comes before~$c'$ along~$f$ in counterclockwise order, starting from~$c_0$.
			There are two routes between~$x$ and~$y$, one along each side of~$f$.
			\begin{enumerate}[label=(\roman*)]
				\item\label{route:1} from~$x$ to~$c$ in~$G_c$, counterclockwise along~$f$ to~$c'$, and then to~$y$ in~$G_{c'}$;
				\item\label{route:2} from~$y$ to~$c'$ in~$G_{c'}$, counterclockwise along~$f$ to~$c$, and then to~$x$ in~$G_c$.
			\end{enumerate}
			We show that one of them is monotone.
			
			The numbers~$u_r$ from~\ref{D:window} satisfy~$u_0 \le u_1 \le \dots \le u_m \le u_0 + 2k$. In the main case, we have~$u_0 = F_1 - k + 1$,~$u_1 = F_1$, and~$u_2 = \dots = u_m = F_1 + k - 1$; in the root case,~$u_r$ is the smallest number that is at least~$F_r$ and belongs to a wedge of~$T$.
			So, as we walk counterclockwise along~$f$ from~$c_0$, the numbers of the edges never decrease; the edges of~$P_0$ get the number~$u_0 + 2k$ at the end of the walk.
			For a vertex~$v$ of~$f$, let~$\mathrm{in}(v)$ and~$\mathrm{out}(v)$ be the numbers of the wedges of the edges entering and leaving~$v$ in this walk, except that~$\mathrm{in}(c_0) = u_0$.
			So~$\mathrm{in}(c_r) = u_r$ and~$\mathrm{out}(c_r) = u_{r+1}$, where~$u_{m+1} = u_0 + 2k$, and~$\mathrm{in}(v) \le \mathrm{out}(v)$ for every vertex~$v$ of~$f$.
			By~\ref{D:window}, and since~$L_r + 1 = F_{r+1} \le u_{r+1}$ for~$r < m$ and~$L_m + 1 = 2k \le u_{m+1}$, we have, for every~$r$,
			\[F_r \le \mathrm{in}(c_r) \le F_r + k - 1 \qquad \text{and} \qquad L_r < \mathrm{out}(c_r).\]
			
			For a vertex~$x \ne c$ of a subgraph~$G_c$, a route that reaches~$x$ along~$f$ via~$c$ first utilizes wedges with numbers up to~$\mathrm{in}(c)$ along~$f$, and then wedges of~$\W_c$ on the path from~$c$ to~$x$, the last of which has the number~$L_c$.
			Similarly, a route that leaves~$x$ via~$c$ and continues along~$f$ first utilizes wedges of~$-\W_c$ on the path from~$x$ to~$c$, the first of which,~$-F_c$, has the number~$F_c + k$, and then wedges with numbers from~$\mathrm{out}(c)$ on along~$f$.
			Thus, we let~$\mathrm{in}(x) = \max\{\mathrm{in}(c), L_c\}$ and~$\mathrm{out}(x) = \min\{\mathrm{out}(c), F_c + k\}$; the first route utilizes wedges with numbers at most~$\mathrm{in}(x)$, and the second one wedges with numbers at least~$\mathrm{out}(x)$.
			Then
			\[\mathrm{in}(c) \le \mathrm{in}(x) \le \mathrm{out}(x) \le \mathrm{out}(c).\]
			The first and the last inequality hold by definition.
			For the middle one, we compare each candidate for~$\mathrm{in}(x)$ with each candidate for~$\mathrm{out}(x)$.
			\begin{itemize}
				\item $\mathrm{in}(c) \le \mathrm{out}(c)$, as for every vertex of~$f$;
				\item $\mathrm{in}(c) < F_c + k$ and~$L_c < \mathrm{out}(c)$, by the two inequalities above;
				\item $L_c < F_c + k$, since~$|\W_c| \le k$.
			\end{itemize}
			Only the first comparison can be an equality, so~$\mathrm{in}(x) = \mathrm{out}(x)$ implies~$\mathrm{in}(c) = \mathrm{out}(c)$.
			
			We now show that one of the routes is short.
			The numbers never decrease as we walk counterclockwise along~$f$ from~$c$ via~$c'$ back to~$c$, so~$\mathrm{out}(c) \le \mathrm{in}(c')$ and~$\mathrm{out}(c') \le \mathrm{in}(c) + 2k$.
			Together with the inequalities above for~$x$ and~$y$, this gives
			\[\mathrm{in}(x) \le \mathrm{out}(x) \le \mathrm{in}(y) \le \mathrm{out}(y) \le \mathrm{in}(x) + 2k.\]
			These four numbers split the~$2k$ steps from~$\mathrm{in}(x)$ to~$\mathrm{in}(x) + 2k$ into four parts, so
			\begin{equation*}
				\begin{split}
					&\bigl(\mathrm{in}(y) - \mathrm{out}(x)\bigr) + \bigl(\mathrm{in}(x) + 2k - \mathrm{out}(y)\bigr)\\
					{}={} &2k - \bigl(\mathrm{out}(x) - \mathrm{in}(x)\bigr) - \bigl(\mathrm{out}(y) - \mathrm{in}(y)\bigr) \le 2k.
				\end{split}
			\end{equation*}
			If both terms on the left were at least~$k$, then both would be exactly~$k$, and we would have~$\mathrm{in}(x) = \mathrm{out}(x)$ and~$\mathrm{in}(y) = \mathrm{out}(y)$.
			This gives~$\mathrm{in}(c) = \mathrm{out}(c)$ and~$\mathrm{in}(x) = \mathrm{in}(c)$.
			Now~$\mathrm{in}(c)$ is the number of the wedge of the edge of~$f$ that enters~$c$, and this edge belongs to a path~$P_r$, so~$\mathrm{in}(x)$ is a number of the wedge~$U_r$.
			Likewise,~$\mathrm{in}(y) = \mathrm{in}(c')$ is a number of the wedge~$U_s$ of the path~$P_s$ that contains the edge entering~$c'$.
			Since~$\mathrm{in}(y) - \mathrm{in}(x) = \mathrm{in}(y) - \mathrm{out}(x) = k$, the wedges~$U_r$ and~$U_s$ are opposite, which contradicts~\ref{D:opp}.
			So~$\mathrm{in}(y) - \mathrm{out}(x) \le k - 1$ or~$\mathrm{in}(x) + 2k - \mathrm{out}(y) \le k - 1$.
			We assume the former; the latter case is symmetric, with the roles of~$x$ and~$y$ exchanged and~$2k$ added to the numbers of~$c$,~$x$, and~$\W_c$.

			We now show that route~\ref{route:1} is monotone.
			We split route~\ref{route:1} at~$c$ into two parts, the part from~$c$ to~$x$ in~$G_c$, and the part from~$c$ along~$f$ to~$c'$ and then to~$y$ in~$G_{c'}$.
			We glue the two parts together with \cref{lem:reach_disjoint}.
			For this, we need two disjoint short sets such that~$x$ is reachable from~$c$ within the first one and~$y$ is reachable from~$c$ within the second one.
			For~$x$, we take~$\W_c$; by assumption,~$x$ is reachable from~$c$ within~$\W_c$.
			For~$y$, we collect the wedges of the second part.
			Its path along~$f$ from~$c$ to~$c'$ uses wedges with numbers from~$\mathrm{out}(c)$ to~$\mathrm{in}(c')$.
			Its path from~$c'$ to~$y$ uses wedges of~$\W_{c'}$, which have the numbers from~$F_{c'}$ to~$L_{c'}$.
			Both ranges lie in the range from~$\min\{\mathrm{out}(c), F_{c'}\}$ to~$\max\{\mathrm{in}(c'), L_{c'}\} = \mathrm{in}(y)$.
			Let~$\mathcal{S}$ be the set of the wedges with numbers in this range.
			(If~$y = c'$, there is no path from~$c'$ to~$y$, and~$\mathcal{S}$ consists of the wedges with numbers from~$\mathrm{out}(c)$ to~$\mathrm{in}(c')$.)
			\begin{itemize}
				\item $y$ is reachable from~$c$ within~$\mathcal{S}$.
				Let~$d \in \D(\mathcal{S})$.
				All wedges of the path along~$f$ from~$c$ to~$c'$ lie in~$\mathcal{S}$, so this path is $d$-monotone.
				Since~$\W_{c'} \subseteq \mathcal{S}$, we have~$d \in \D(\W_{c'})$, so the path can be continued by a $d$-monotone path from~$c'$ to~$y$.
				\item $\mathcal{S}$ is short, that is,~$\mathrm{in}(y)$ is at most~$k-1$ larger than both~$\mathrm{out}(c)$ and~$F_{c'}$.
				First,~$\mathrm{in}(y) - \mathrm{out}(c) \le \mathrm{in}(y) - \mathrm{out}(x) \le k - 1$ by our assumption.
				Second,~$\mathrm{in}(y)$ is~$\mathrm{in}(c')$ or~$L_{c'}$, and both are at most~$F_{c'} + k - 1$, by~\ref{D:window} and since~$|\W_{c'}| \le k$.
				\item $\mathcal{S}$ and~$\W_c$ are disjoint.
				The wedges of~$\W_c$ have the numbers from~$F_c$ to~$L_c$ and, one period later, from~$F_c + 2k$ to~$L_c + 2k$.
				All numbers of~$\mathcal{S}$ lie strictly between~$L_c$ and~$F_c + 2k$.
				Indeed,~$\mathrm{out}(c)$ and~$F_{c'}$ are larger than~$L_c$, because the wedge sets after~$\W_c$ start at~$L_c + 1$; and~$\mathrm{in}(y) \le \mathrm{out}(x) + k - 1 \le F_c + 2k - 1$, by our assumption and since~$\mathrm{out}(x) \le F_c + k$.
			\end{itemize}
			Now \cref{lem:reach_disjoint} yields a monotone path between~$x$ and~$y$.
			(If~$x = c$, we do not need the lemma, since~$\D(\mathcal{S}) \ne \emptyset$ and, hence, the first item already gives a monotone path from~$c$ to~$y$.)
			This shows~(ii).

			For (i), let~$d \in \D(\mathcal{A})$.
			We argue as in the proof of \cref{lem:face-one-wedge}. The interior angle at~$c_0$ lies in~$\mathcal{A}$, so both edges of~$f$ at~$c_0$ make positive progress in direction~$d$, and~$c_0$ is the unique lowest vertex of~$f$ in direction~$d$.
			Since~$f$ is strictly convex, every vertex of~$f$ is reached from~$c_0$ by a $d$-monotone path along~$f$.
			For~$x \in G_{c_r} - c_r$, we have~$\W_{c_r} \subseteq \mathcal{A}$, so~$d \in \D(\W_{c_r})$, and we append a $d$-monotone path from~$c_r$ to~$x$.

	\end{proof}
	
	\threeLeaves*
	\label{lem:three-leaves*}
	\begin{proof}
		Let~$f$ be a high-degree face of~$G$, and let~$v_f$ be the corresponding node of~$T_G$, which has degree at least three.
		Since~$T_G$ has only three leaves,~$v_f$ has degree exactly three, and each of the three components of~$T_G - v_f$ contains exactly one leaf, so it is a path.
		Hence,~$f$ is the only high-degree face, and it has exactly three cut vertices~$c_0$, $c_1$, and~$c_2$, which we name in clockwise order along~$f$ in~$\mathcal E$; the subgraphs that hang off them are maximal chains~$M_0$, $M_1$, and~$M_2$, respectively, each of which ends in a leaf of~$T_G$.
		Let~$\langle c_0, a_1, \dots, a_q, c_1 \rangle$, $\langle c_1, b_1, \dots, b_r, c_2 \rangle$, and~$\langle c_2, w_1, \dots, w_s, c_0 \rangle$ be the three boundary paths of~$f$, where~$q, r, s \ge 0$, and let~$w_s = c_2$ if~$s = 0$.
		As in the proof of \cref{thm:lowerbound}, we may assume, after an orientation-preserving affine transformation, that the two lines of~\LL are the x- and the y-axis, so the four wedges of~\WL are the quadrants.

		Let~$P$ be the path that consists of~$c_0$, $a_1$, \dots, $a_q$, $c_1$, $b_1$, \dots, $b_r$, $c_2$, $w_1$, \dots, $w_s$. We place its vertices in this order from left to right on the upper half of a circle such that~$c_1$ is its topmost point; see~\cref{fig:high_degree_face_3}.
		Then~$P$ is x-monotone, its edges up to~$c_1$ point to the northeast, its remaining edges point to the southeast, and~$P$ together with the edge~$(w_s, c_0)$ bounds a convex polygon whose boundary is traversed clockwise in the order~$c_0$, $c_1$, $c_2$, so the drawing of~$f$ respects~$\mathcal E$.
		We draw~$M_0$, $M_1$, and~$M_2$ by \cref{lem:chain} inside the southwest quadrant at~$c_0$, the northwest quadrant at~$c_1$, and the northeast quadrant at~$c_2$, respectively.
		Hence, the drawing is planar, and its bounded faces are convex.

		For~$i \in \{0, 1, 2\}$, every vertex of~$M_i$ is reached from~$c_i$ by a path in~$M_i$ that utilizes only the quadrant of~$M_i$ by \cref{lem:chain}, so this path is both x- and y-monotone; moreover, any two vertices of~$M_i$ are connected by an \LL-monotone path.
		Any two vertices of~$f$ are connected along the x-monotone path~$P$.
		Now let~$u$ be a vertex of~$M_i$, and let~$v$ be a vertex of~$f$ or of~$M_j$ with~$j \ne i$.
		We connect~$u$ and~$v$ by the path that goes from~$u$ to~$c_i$ in~$M_i$, then along~$P$, and, if~$v$ lies on~$M_j$, from~$c_j$ to~$v$ in~$M_j$.
		If~$i = 0$, this path is x-increasing if~$v$ lies on~$f$ or on~$M_2$, and it is y-increasing if~$v$ lies on~$M_1$.
		If~$i = 1$, it is y-decreasing if~$v$ lies on~$f$, and it is x-increasing if~$v$ lies on~$M_2$.
		If~$i = 2$ and~$v$ lies on~$f$, it is x-decreasing if~$v$ precedes~$c_2$ on~$P$, and it is y-decreasing otherwise.
		The remaining pairs are covered by reversing these paths.

	\end{proof}
	
	\upperBound*
	\label{thm:upperbound*}
	\begin{proof}
                  As mentioned at the start of \cref{sec:cacti}, we
                  first show that it suffices to prove the statement
                  for cacti.  To this end, we delete from~$G$ all chords.
		This yields a spanning cactus~$G'$ of~$G$ with~$T_{G'} = T_G$ and~$\ell(G') = \ell(G)$; moreover, every block of~$G'$ contains at most two cut vertices if and only if this holds for~$G$, that is, if and only if~$G'$ has no high-degree face.
		Suppose that~$G'$ has a drawing as claimed, with the embedding that~$\mathcal E$ induces on~$G'$.
		Every deleted chord connects two vertices on a bounded face of~$G'$, and we reinsert it as a straight-line segment inside this face.
		Since the faces are convex and the deleted chords of a block do not cross in~$\mathcal E$, the drawing remains planar, its new bounded faces are also convex, and it realizes~$\mathcal E$.
		Since adding edges preserves monotone paths, the drawing remains \LL-monotone.
		Hence, we assume from now on that~$G$ is a cactus.
		\Cref{lem:high_degree_faces} requires at least three lines, so we first consider the case that~$G$ has a high-degree face and~$|\LL| = 2$. By assumption,~$\lceil \ell(G)/2 \rceil \le |\LL| = 2$, so~$\ell(G) \le 4$, and~$\ell(G) \ne 4$, since otherwise~\LL would contain at least three lines. Moreover, the node of a high-degree face has degree at least three in~$T_G$, so~$T_G$ has at least three leaves. Hence,~$\ell(G) = 3$, and \cref{lem:three-leaves} yields the desired drawing. In the remainder of the proof, we can therefore assume that~$|\LL| \ge 3$ whenever~$G$ has a high-degree face.

		Let~$k=|\LL|$ and let~$\WL$ denote the set of~$2k \ge \ell(G)$ wedges induced by the lines in~$\overline{\LL}$.
		If~$G$ consists of a single block, that is, of a single vertex, a single edge, or a cycle, then the statement is trivial in the first two cases, and, for a cycle, \cref{lem:face-1-wedge} with any vertex as~$c$ and any single wedge as~\W yields a convex \LL-monotone drawing. Hence, we assume that~$G$ has a cut vertex, so~$T_G$ has at least three nodes.
		We assign wedges to the leaves of~$T_G$ following a strategy similar to the one used in \cref{thm:trees}.
		Let~$v^*$ be a node of~$T_G$ that is not a leaf such that every component of~$T_G - v^*$ contains at most~$\ell(G)/2$ leaves of~$T_G$. If~$\ell(G) \ge 3$, we can take a leaf centroid of~$T_G$. If~$\ell(G) = 2$, then~$T_G$ is a path with at least three nodes, and we can take any of its inner nodes.
		If~$v^*$ corresponds to a cut vertex in $G$, let~$c^\star$ be this cut vertex; otherwise, let~$c^\star$ be a cut vertex of the block corresponding to~$v^*$.		
		Root~$T_G$ at~$v_{c^\star}$ and enumerate its leaves as~$\lambda_0, \dots, \lambda_{\ell(G)-1}$ in the order that they are encountered in a DFS traversal of~$T_G$. As in the proof of \cref{thm:trees}, the DFS visits the children of every node in the order given by~$\mathcal E$, starting, at every node other than the root, with the child that follows the parent in this order, and the wedges of~$\WL$ are numbered in counterclockwise order; hence, the subtrees of every node receive consecutive wedges in the cyclic order of~$\mathcal E$, so the drawing respects~$\mathcal E$.
		We assign to the leaves~$\lambda_0, \dots, \lambda_{\ell(G)-1}$, in this order, non-empty sets~$\mathcal{A}_{\lambda_i}$ of consecutive wedges of~\WL, such that every wedge is assigned to exactly one leaf and the leaves of every component of~$T_G - v^*$ receive at most~$k$ wedges in total.
		For this, we give every leaf one wedge and distribute the remaining~$2k - \ell(G)$ wedges among the leaves. This is possible since~$v^*$ is not a leaf, so~$T_G - v^*$ has at least two components, and each of them has at most~$\ell(G)/2 \le k$ leaves.
		For every vertex~$w$ of~$T_G$, let~$\mathcal{A}_{w}$ denote the wedge set assigned to the leaves of the subtree~$T_{w}$ rooted at~$w$.
		Note that this notation is consistent in case~$w$ is a leaf.
		For every node~$w \notin \{v_{c^\star}, v^*\}$, the subtree~$T_w$ lies in a component of~$T_G - v^*$; hence~$|\mathcal{A}_w| \le k$.
		
		As outlined at the beginning of \cref{sec:cacti}, the construction of the drawing for~$G$ follows the proof of \cref{thm:trees}, applied to~$T_G$. For an \LL-monotone drawing of a cactus $G$ refer to \cref{fi:bc_tree_2}.
		However, instead of embedding leaf paths and branches, we embed maximal chains and high-degree faces of~$G$.
		Let~$c$ be \textit{any} cut vertex of~$G$, and let~$v_{c}$ be the corresponding node in~$T_G$. Throughout the construction, we maintain the following invariants.
		\begin{enumerate}[label=\textbf{(I\arabic*)},leftmargin=*,ref=(I\arabic*)]
			\item \label{inv:root-path} Let~$G^{(i)}$ be a component of~$G_c - c$, together with~$c$, and let~$\mathcal{A}^{(i)}$ be the wedge set assigned to the leaves of the corresponding subtree of~$T_G$. Each~$u \in V(G^{(i)})$ is reachable from~$c$ within~$\mathcal{A}^{(i)}$.
			\item \label{inv:monotone-path} Every pair of vertices in~$G_c$ is connected by an~$\LL$-monotone path.
			\item \label{inv:planar} The drawing of~$G_c$ is planar, its bounded faces are convex, and, for each component~$G^{(i)}$ as in~\ref{inv:root-path},~$G^{(i)} - c$ lies in the interior of the region~$\mathcal{A}^{(i)}(c)$.
		\end{enumerate}
		Since the wedge sets of the components of~$G_c - c$ are disjoint,~\ref{inv:planar} implies that these components lie in pairwise disjoint regions; this yields the planarity of the drawing at~$c$.
		If~$v^*$ is a high-degree face, we do not need the invariants at~$c^\star$ for the component that contains~$v^*$, since \cref{lem:high_degree_faces} then directly yields the drawing of~$G$.
		We construct the drawing of~$G$ by traversing~$T_G$
		using a bottom-up approach.  We distinguish two cases
		based on the components that we encounter.
		
		\medskip
		
		\noindent\textbf{Maximal Chains.} Let~$P = \langle v_0, v_1, \dots, v_m \rangle$ be an oriented maximal path of~$T_G$, let~$M_P$ be the maximal chain of~$G$ corresponding to~$P$ and let~$c_0$ be the cut vertex of~$G$ that corresponds to~$v_0$ in~$T_G$.
		If~$v_m$ is a leaf~$\lambda_i$ of~$T_G$, then~$1 \le |\mathcal{A}_{v_m}| \le k$. Otherwise,~$T_{v_m}$ has at least two leaves and~$v_m \notin \{v_{c^\star}, v^*\}$, so~$2 \le |\mathcal{A}_{v_m}| \le k$.
		In both cases, we draw~$M_P$ by \cref{lem:chain} with~$\W = \mathcal{A}_{v_m}$; since~$k \ge 2$, the lemma applies. If~$v_m$ is not a leaf, the inductively computed drawing of~$G_{c_m}$ is translated such that its apex lies at the terminal vertex~$c_m$. Translation trivially preserves the internal monotonicity of paths.

		By \cref{lem:chain}(i), every vertex of~$M_P$ is reachable from~$c_0$ within~$\mathcal{A}_{v_m}$.
		If~$v_m$ is not a leaf, then, by the induction hypothesis and since~$|\mathcal{A}_{v_m}| \le k$, every vertex~$u \in G_{c_m}$ is reachable from~$c_m$ within~$\mathcal{A}_{v_m}$ (reachability within a subset of~$\mathcal{A}_{v_m}$ implies reachability within~$\mathcal{A}_{v_m}$). Hence, for every~$d \in \D(\mathcal{A}_{v_m})$, a $d$-monotone path from~$c_0$ to~$c_m$ followed by a $d$-monotone path from~$c_m$ to~$u$ is $d$-monotone; that is,~$u$ is reachable from~$c_0$ within~$\mathcal{A}_{v_m}$. Since~$M_P$ contains no branching,~$\mathcal{A}_{v_m}$ is the wedge set~$\mathcal{A}^{(i)}$ of the component of~$G_{c_0} - c_0$ that contains~$M_P$, and~$|\mathcal{A}_{v_m}| \le k$, which satisfies invariant~\ref{inv:root-path}.
		The same argument, applied to the subchain of~$M_P$ that starts at an intermediate cut vertex~$x_s$ of~$M_P$ (which is drawn in the same way), shows that~\ref{inv:root-path} also holds at~$x_s$.
		Moreover, by \cref{lem:chain}(iii), there exists an~$\LL$-monotone path between any two vertices in~$M_P$, and, by the induction hypothesis, between any two vertices in~$G_{c_m}$.
		For a vertex~$w \in M_P$ and a vertex~$u \in G_{c_m}$, let~$d \in \D(\mathcal{A}_{v_m})$, which is non-empty since~$|\mathcal{A}_{v_m}| \le k$. By \cref{lem:chain}(ii), there is a $d$-monotone path from~$w$ to~$c_m$, and there is a $d$-monotone path from~$c_m$ to~$u$. Their concatenation at~$c_m$ is $d$-monotone, satisfying invariant~\ref{inv:monotone-path} for pairs of vertices of~$M_P \cup G_{c_m}$.
		
		For~\ref{inv:planar}, the chain~$M_P$ is planar and lies inside~$\mathcal{A}_{v_m}(c_0)$ by \cref{lem:chain}; its faces are convex by \cref{lem:face-1-wedge,lem:face-2-wedges,lem:face-one-wedge}.
		If~$v_m$ is not a leaf, then~$G_{c_m} - c_m$ lies in the interior of~$\mathcal{A}_{v_m}(c_m)$ by the induction hypothesis, and~$c_m$ lies in~$\mathcal{A}_{v_m}(c_0)$, so~$G_{c_m}$ lies inside~$\mathcal{A}_{v_m}(c_0)$, which is convex as~$|\mathcal{A}_{v_m}| \le k$.
		Moreover, every vertex of~$M_P$ other than~$c_m$ reaches~$c_m$ by a path whose edges point into the interiors of wedges of~$\mathcal{A}_{v_m}$ (see the proof of \cref{lem:chain}), so~$M_P - c_m$ lies in the interior of the reflection of~$\mathcal{A}_{v_m}(c_m)$ at~$c_m$; since~$\mathcal{A}_{v_m}(c_m)$ is convex, this interior is disjoint from~$G_{c_m} - c_m$.
		Hence, the drawing of~$M_P \cup G_{c_m}$ is planar.
		
		\medskip
		
		\noindent\textbf{High-Degree Faces.} Let~$f$ be a high-degree face of~$G$ with node~$v_f$ in~$T_G$, parent vertex~$c_0$, and child vertices~$c_1, \dots, c_m$ in counterclockwise order.
		We draw~$f$ together with~$G_{c_1}, \dots, G_{c_m}$ by \cref{lem:high_degree_faces} with~$\W_{c_r} = \mathcal{A}_{v_{c_r}}$ for~$1 \le r \le m$ and~$\W_{c_0} = \WL \setminus \mathcal{A}_{v_f}$.
		These sets appear in counterclockwise order, since the DFS follows~$\mathcal E$; we have~$|\W_{c_r}| \le k$ for~$r \ge 1$, since~$v_{c_r} \notin \{v_{c^\star}, v^*\}$; $\W_{c_0} \ne \emptyset$, since~$c^\star$ lies in at least two blocks; and~$k \ge 3$, since~$G$ has a high-degree face.
		The assumptions of the lemma on~$G_{c_1}, \dots, G_{c_m}$ are~\ref{inv:root-path}--\ref{inv:planar} at~$c_1, \dots, c_m$.
		
		If~$f \ne v^*$, then~$|\mathcal{A}_{v_f}| \le k$, and the lemma yields the invariants for the component~$f \cup G_{c_1} \cup \dots \cup G_{c_m}$ of~$G_{c_0} - c_0$.
		If~$f = v^*$, then~$c_0 = c^\star$, and~$\W_{c_0}$ consists of the wedges of the other components of~$G_{c^\star} - c^\star$, which form one component of~$T_G - v^*$; so~$|\W_{c_0}| \le k$.
		We draw these other components first and take their union as~$G_{c_0}$; by the invariants and the next paragraph, it satisfies the assumptions of the lemma, which then yields a planar \LL-monotone drawing of~$G$.
				
		It remains to show~\ref{inv:monotone-path} for pairs of vertices in different components of~$G_c - c$; in the two cases above, we have shown it for pairs within a component.
		Let~$c$ be a cut-vertex of~$G$ with corresponding node~$v_c \in T_G$, and let~$G^{(1)}, \dots, G^{(q)}$ be the connected components of~$G_c$ rooted at~$c$, corresponding to subtrees~$T^{(1)}, \dots, T^{(q)}$ of~$T_G$. To show an~$\LL$-monotone path between any~$u \in G^{(i)}$ and~$w \in G^{(j)}$ ($i \ne j$), let~$\mathcal{A}^{(i)}$ and~$\mathcal{A}^{(j)}$ be the disjoint wedge sets assigned to the leaves of~$T^{(i)}$ and~$T^{(j)}$. By Invariant~\ref{inv:root-path},~$u$ and~$w$ are reachable from~$c$ within sets of at most~$|\LL|$ consecutive wedges of~$\mathcal{A}^{(i)}$ and~$\mathcal{A}^{(j)}$, respectively. By \cref{lem:reach_disjoint},~$u$ and~$w$ are connected by an~$\LL$-monotone path.

	\end{proof}
	
	\lowerBound*
	\label{thm:lowerbound*}
	\begin{proof}
          We first show~(i). For~$\ell \ge 3$, let~$G_\ell$ be a star
          with~$\ell$ leaves. Then~$\ell(G_\ell)=\ell$,
          and~$|\LL| \ge \lceil \ell/2 \rceil$ by \cref{thm:trees}.
		
          For~$\ell = 2$, let~$G_2$ consist of a
          $4$-cycle~$\langle s,a,t,b \rangle$ and two pendant
          edges~$(s,u)$ and~$(t,v)$.  Clearly,~$\ell(G_2)=2$.
          Suppose, for a contradiction, that~$G_2$ admits an
          \LL-monotone drawing with $|\LL|=1$.  Let $\LL=\{l\}$.
          Rotate the drawing so that~$l$ is vertical.  Then a path is
          \LL-monotone if and only if it is y-monotone.
		
          Every path from~$u$ to~$v$ passes through~$s$ and~$t$, so we
          may assume that~$u$,~$s$,~$t$, and~$v$ have increasing
          y-coordinates.  Every path from~$a$ to~$b$ passes
          through~$s$ or~$t$.  By symmetry, we may assume that the
          path~$\langle a,s,b \rangle$ is y-monotone, and (after
          exchanging~$a$ and~$b$ if necessary) that~$a$ lies
          below~$s$.  Then no path from~$a$ to~$u$ is y-monotone since
          every such path passes through~$s$, which lies above
          both~$a$ and~$u$.  This is the desired contradiction.
		
          Now we show~(ii).  Let~$G$ be the cactus that consists of
          the cycle~$C = \langle v_0, v_1, \dots, v_{11} \rangle$ and
          four pendant edges at~$v_0$, $v_3$, $v_6$, and~$v_9$.
          Clearly,~$\ell(G) = 4$.  Suppose, for a contradiction,
          that~$G$ admits an \LL-monotone drawing with~$|\LL| = 2$.
          Let $\{l_0, l_1\} = \LL$.
          By applying an affine transformation to the plane, which
          preserves planarity and monotonicity, we may assume
          that~$l_0$ and~$l_1$ are the x- and the y-axis.
          Then a path is \LL-monotone if and only if it is x-monotone
          or y-monotone.  We may assume that, in the transformed
          drawing, no two vertices have the same x- or y-coordinates.
		
		If two pendant edges, directed away from~$C$, pointed into the same quadrant, then the path between their leaves would start and end with edges pointing into opposite quadrants, so it would be neither x- nor y-monotone.
		Hence, the four pendant edges point into the four quadrants; let~$p'$, $q'$, $r'$, and~$s'$ be the leaves whose pendant edges point to the NE, NW, SW, and SE, respectively, and let~$p$, $q$, $r$, and~$s$ be the vertices of~$C$ that carry them.
		Any two of these vertices have distance at least~$3$ along~$C$.
		Every path from~$p'$ starts with an edge pointing to the SW, so, if it is \LL-monotone, it is x-decreasing or y-decreasing; similarly, every \LL-monotone path from~$q'$ is x-increasing or y-decreasing, from~$r'$ it is x-increasing or y-increasing, and from~$s'$ it is x-decreasing or y-increasing.
		
		Let~$\lambda$, $\tau$, $\rho$, and~$\beta$ be the leftmost, topmost, rightmost, and bottommost vertex of~$C$, respectively.
		They are vertices of the convex hull of~$C$ and appear along the hull in the counterclockwise order~$\rho$, $\tau$, $\lambda$, $\beta$.
                Since~$C$ is a simple polygon, it visits these vertices
                in the same order (if we orient~$C$ counterclockwise).
		Two consecutive vertices among the four may
                coincide, but~$\lambda \ne \rho$ and~$\tau \ne \beta$.
		Let~$P_1$, $P_2$, $P_3$, and~$P_4$ be the subpaths of~$C$ from~$\rho$ to~$\tau$, from~$\tau$ to~$\lambda$, from~$\lambda$ to~$\beta$, and from~$\beta$ to~$\rho$, respectively, in counterclockwise direction; such a path consists of a single vertex if its endpoints coincide.
		We use the following observation, which holds since~$\lambda$, $\rho$, $\beta$, and~$\tau$ have the smallest and largest x- and y-coordinates on~$C$.
		\begin{enumerate}[label=($\star$),leftmargin=*]
                \item\label{enum:subpath} Let~$Q$
                  be a subpath of~$C$.  If~$Q$ is x-increasing,
                  then~$\lambda$ can only be its first vertex and~$\rho$ can
                  only be its last vertex; if~$Q$ is x-decreasing,
                  then~$\rho$ can only be its first vertex and~$\lambda$ can
                  only be its last vertex. The same holds for
                  y-monotone paths with~$\beta$ and~$\tau$ instead of~$\lambda$
                  and~$\rho$.
		\end{enumerate}
		
		For a path that starts in one of the leaves~$p'$, $q'$, $r'$, and~$s'$, we apply~\ref{enum:subpath} to its subpath that starts at the vertex of~$C$ that carries this leaf.
		
		Suppose that $P_1$ contains an inner vertex~$u$,
                and that $P_3$ contains an inner vertex~$w$.
		One of the two paths between~$u$ and~$w$ contains~$\tau$ and~$\lambda$ as inner vertices;
                the other one contains~$\beta$ and~$\rho$ as inner vertices.
		By~\ref{enum:subpath}, neither of them is x-monotone or y-monotone,
                so~$u$ and~$w$ are not connected by an \LL-monotone path.
		Hence,~$P_1$ or~$P_3$ has no inner vertex.
                By the same argument,~$P_2$ or~$P_4$ has no inner vertex.
		We consider two cases.
                
        First, we assume that $\lambda$, $\tau$, $\rho$, and~$\beta$ are pairwise distinct; see~\cref{fig:deg_four_1}.
		We show that~$p$ lies on~$P_1$. The path from~$p'$ to~$\tau$ is x-decreasing.
		Otherwise, it is y-decreasing, so~$p = \tau$ by~\ref{enum:subpath}, and then the
		path is the edge~$(p',\tau)$, which points to the SW and, hence, is x-decreasing
		as well. By~\ref{enum:subpath}, the path avoids~$\lambda$ and contains~$\rho$ only
		if~$p = \rho$; hence,~$p \in P_1 \cup P_2$ and~$p \ne \lambda$.
		Similarly, the path from~$p'$ to~$\rho$ is y-decreasing, so~$p \in P_4 \cup P_1$ and~$p \ne \beta$; as~$P_2$ and~$P_4$ are disjoint,~$p \in P_1$.
		Symmetrically,~$q$ lies on~$P_2$, $r$ on~$P_3$, and~$s$ on~$P_4$.
		By the previous paragraph, two consecutive paths among~$P_1, \dots, P_4$ have no inner vertices, say~$P_1$ and~$P_2$; together, they have at most two edges.
		Then~$p$ and~$q$ have distance at most~$2$ along~$C$, a contradiction.
		
		Second, we assume that two of~$\lambda$, $\tau$, $\rho$, and~$\beta$ coincide; see~\cref{fig:deg_four_2}.
		By symmetry, let~$\tau = \lambda$, so that~$P_2$ consists of a single vertex;
		then~$\tau = \lambda = q$.  Otherwise, the path from~$q'$ to~$\tau, \lambda$
                would end with an edge of~$C$, so, by~\ref{enum:subpath},
                it could be neither x-increasing nor y-decreasing.
		As shown above,~$P_1$ or~$P_3$ has no inner vertex;
                by symmetry, we may assume that~$P_1$ has no inner vertex,
                that is,~$P_1$ is the edge~$(\rho, \tau)$.
		As~$\rho$ is adjacent to~$q$, it carries no pendant edge.
		By~\ref{enum:subpath}, the path from~$p'$ to~$\rho$ cannot be x-decreasing,
                so it is y-decreasing; similarly, the path from~$s'$ to~$\rho$ is y-increasing.
		By~\ref{enum:subpath}, neither of the two paths contains~$\tau$
                since the first one would have to start there
                and the second one would have to end there.
		Hence, both paths enter~$\rho$ along the same edge,
                which would have to point down and up at the same time,
                a contradiction.
        \begin{figure}[tb]
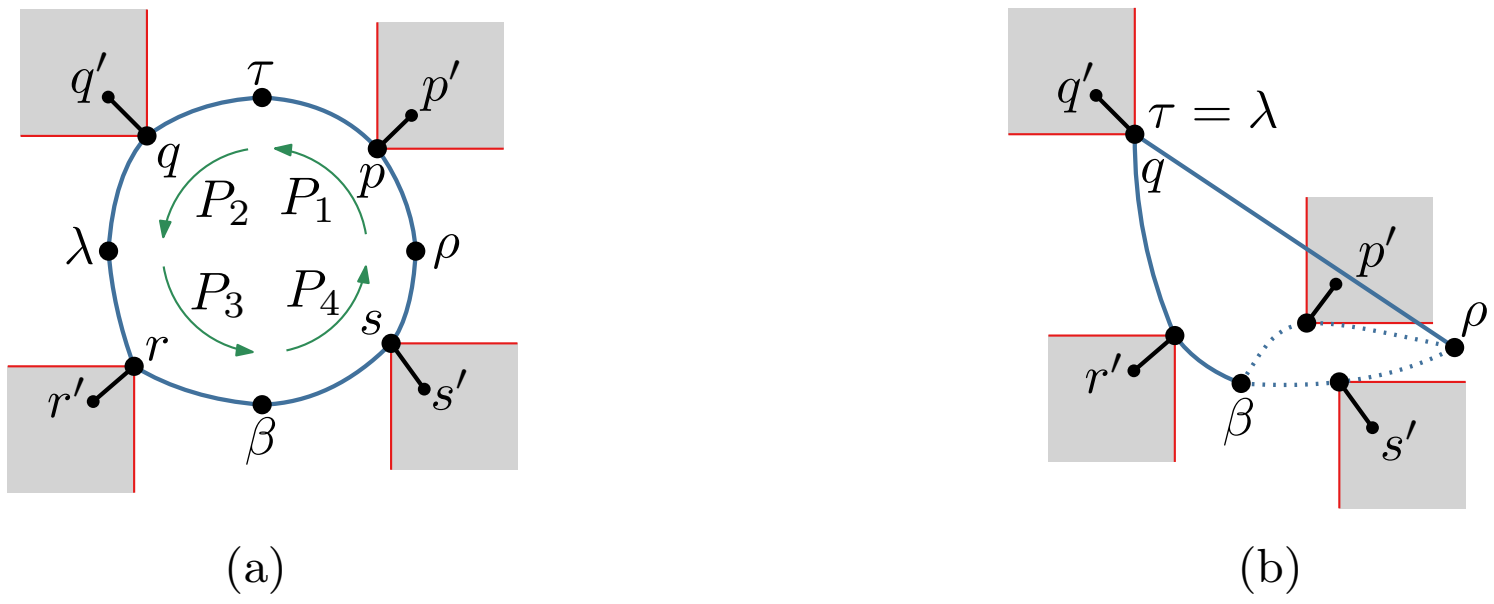

              	\captionsetup[subfigure]{justification=centering}
              	\begin{subfigure}{.45\linewidth}
              		\centering
              		\includegraphics[page=4]{high_degree_faces}
              		\subcaption{}
              		\label{fig:deg_four_1}
              	\end{subfigure}
              	\hfill
              	\begin{subfigure}{.45\linewidth}
              		\centering
              		\includegraphics[page=5]{high_degree_faces}
              		\subcaption{}
              		\label{fig:deg_four_2}
              	\end{subfigure}
              	\caption{The two cases in the proof of \cref{thm:lowerbound}(ii).
              		(a)~The extreme vertices~$\lambda$, $\tau$, $\rho$, and~$\beta$ are pairwise distinct; then the vertices~$p$, $q$, $r$, and~$s$ lie on~$P_1$, $P_2$, $P_3$, and~$P_4$, respectively.
              		(b)~Two extreme vertices coincide, here~$\tau = \lambda = q$, and~$P_1$ is the edge~$(\rho, \tau)$; the paths from~$p'$ and from~$s'$ to~$\rho$ (dotted) enter~$\rho$ along the same edge, which would have to point both down and up.}
              	\label{fig:deg_four}
        \end{figure}
	\end{proof}

\end{document}